\documentclass[reprint,amsmath,amssymb,aps]{revtex4-2}
\usepackage{natbib}
\usepackage{graphicx}
\usepackage{dcolumn}
\usepackage{bm}
\usepackage{textcase}
\usepackage{url}
\usepackage[hidelinks]{hyperref}
\usepackage[mathlines]{lineno}
\usepackage{float}
\begin{document}
\preprint{APS/123-QED}

\title{System size dependence of thermodynamic variables at kinetic freeze-out in high-energy collisions using the Tsallis distribution}

\author{Rishabh Sharma}
\email{rishabhsharma@students.iisertirupati.ac.in}
\author{Krishan Gopal}
\email{krishangopal@students.iisertirupati.ac.in}
\author{Sharang Rav Sharma}
\email{sharang.rav@students.iisertirupati.ac.in}
\author{Chitrasen Jena}
 \email{cjena@iisertirupati.ac.in}
\affiliation{%
 Department of Physics, Indian Institute of Science Education and Research (IISER) Tirupati, Tirupati 517507, Andhra Pradesh, India
}%


\begin{abstract}
We use a thermodynamically consistent form of Tsallis distribution to study the dependence of various thermodynamic quantities on the system size in high-energy collisions. The charged hadron spectra obtained in $p$+$p$, $p$+Pb, Xe+Xe, and Pb+Pb collisions at LHC are used to determine the energy density, pressure, particle density, entropy density, mean free path, Knudsen number, heat capacity, isothermal compressibility, expansion coefficient, and speed of sound at the kinetic freeze-out surface. These quantities are studied as a function of the system size. Notably, the rate of increase (or decrease) in these thermodynamic variables is found to be more rapid in small systems such as $p$+$p$ and $p$+Pb collisions than in large systems such as Xe+Xe and Pb+Pb collisions. This may be due to the small volume of the hadronic system in small collision systems at kinetic freeze-out. It is observed that high-multiplicity $p$+$p$ collisions produce similar thermodynamic conditions as peripheral heavy-ion collisions at kinetic freeze-out. 
\end{abstract}

\keywords{Suggested keywords}
\maketitle

\section{\label{sec:intro}Introduction}
High-energy collisions provide a unique opportunity to study the behavior of matter at extraordinarily high temperatures and densities. In the last few decades, considerable progress has been made in the study of the properties of a new state of matter made up of de-confined quarks and gluons called Quark-Gluon Plasma (QGP). This state is believed to have formed when normal baryonic matter undergoes a first-order or crossover phase transition at very high temperature and/or high baryon density \cite{PhysRevD.78.074507,Aoki2006,Adams2005}. Previously, it was believed that QGP is only formed in heavy-ion collisions; however, recent results show that high-multiplicity $p$+$p$ collisions behave similarly to heavy-ion collisions, posing the question of whether QGP droplets are formed in such collisions \cite{2017,Bjorken2013,Khachatryan2017,Khuntia2019}. Consequently, there is a pressing need to gain deeper insights into high-multiplicity $p$+$p$ collisions in order to understand the properties of the produced system.

To better understand the behavior of matter resulting from high-energy collisions, it is crucial to comprehensively understand its thermodynamic properties \cite{Azmi2020,Deb2021,Sahu2021,Khuntia2016,Jain2023,PhysRevC.100.014910,PhysRevC.96.044901,https://doi.org/10.48550/arxiv.1905.06532}. In this study, we have explored various thermodynamic properties of the hadronic medium produced in both small collision systems ($p$+$p$ and $p$+Pb) and large collision systems (Xe+Xe and Pb+Pb) at LHC energies. 

We examine a scenario in which the fireball resulting from high-energy collisions cools down as it undergoes spacetime expansion. The final-state particles decouple from the system after Kinetic Freeze-Out (KFO), which represents the surface of the last elastic scattering of the produced hadrons. The production of final-state particles in abundance warrants the use of statistical methods to describe such systems. One such approach involves studying the transverse momentum ($p_{T}$) spectra of the produced particles in the framework of Tsallis statistics. Tsallis distribution was initially introduced as a generalization of the Boltzmann-Gibbs distribution and it accounts for deviations from the standard equilibrium statistical description \cite{Tsallis1988}. Over time, Tsallis distribution has proven to be a valuable tool for describing the hadron spectra in high-energy collisions. The Tsallis distribution function is defined as \cite{Cleymans2012}:

\begin{equation}
\label{eq:tsa_dist}
f(E,q,T,\mu) = \left[1+(q-1)\frac{E-\mu}{T}\right]^{\frac{-1}{q-1}},
\end{equation}

where $E$ is the energy of the particle, $q$ is the entropy index, $T$ is the temperature, and $\mu$ is the chemical potential. $q$ is indicative of the degree of deviation from equilibrium, with $q=1$ representing complete equilibrium. Although various forms of the Tsallis distribution have been used to study the hadron spectra, ensuring thermodynamic consistency in their formulations poses a significan challenge \cite{Hui2018,Br2017,Si2018,Lao2017}. In the present study, we have employed Eq. (\ref{eq:tsa_dist}), which has been demonstrated to maintain thermodynamic consistency \cite{Cleymans2012}. The thermodynamic quantities of interest were derived using Eq. (\ref{eq:tsa_dist}) in accordance with the following relations \cite{Cleymans20122,PhysRevD.94.094026,Bhattacharyya2017}:
\begin{align}
\text{Entropy: }S &= sV = -gV\int\frac{d^3p}{(2\pi)^3}\left(\frac{f-f^q}{1-q}-f\right), \\
\text{Number: }N &= nV = gV\int\frac{d^3p}{(2\pi)^3}f^q, \\
\text{Energy: }E &= \epsilon V = gV\int\frac{d^3p}{(2\pi)^3}Ef^q, \\
\text{Pressure: }P &= g\int\frac{d^3p}{(2\pi)^3}\frac{p^2}{3E}f^q,
\end{align}
where $g$ is the degeneracy factor, $V$ is the Tsallis volume; and $s$, $n$, and $\epsilon$ correspond to entropy density, number density, and energy density, respectively. Tsallis volume, $V$, should not be identified with the volume determined from femtoscopy \cite{physics2040038}. 
Additionally, we have also studied thermodynamic response functions, such as heat capacity ($C_{V}$), isothermal compressibility ($\kappa_{T}$), and expansion coefficient ($\alpha$), along with mean free path ($\lambda$), Knudsen number ($Kn$), and squared speed of sound ($c_{s}^{2}$) in the produced hadronic medium. The primary objective of this work is to understand the interplay between collision energy and system size on various thermodynamic quantities, contrasting their behavior in small and large systems.

The remainder of this paper is organized as follows: In Section \ref{sec:spectra}, we discuss the Tsallis distribution that is used to fit the transverse momentum spectra of the charged hadrons. In Section \ref{sec:thermo}, we calculate the various thermodynamic quantities and discuss their dependence on charged particle multiplicity. Finally, in Section \ref{sec:summary}, we summarize our findings.

\section{\label{sec:spectra}Transverse momentum spectra}

The $p_T$-spectra of different particle species can be accurately described using the Tsallis distribution. In this framework, the invariant yield of a particle is defined as:

\begin{equation}
E\frac{d^3N}{d^3p} = \frac{gV}{(2\pi)^3}E\left[1+(q-1)\frac{E-\mu}{T}\right]^{\frac{-q}{q-1}}.
\end{equation}

This expression can be represented $p_{T}$, transverse mass ($m_{T}=\sqrt{p_T^2+m^2}$, where $m$ is the particle mass), and rapidity ($y$) as follows:

\begin{eqnarray}
\frac{d^2N}{p_Tdp_Tdy} = \frac{gV}{(2\pi)^2}m_T\cosh(y) \nonumber\\\
\times \left[1+(q-1)\frac{m_T\cosh(y)-\mu}{T}\right]^{\frac{-q}{q-1}}.
\end{eqnarray}

Assuming $\mu=0$ at LHC energies, the expression for the invariant yield at mid-rapidity ($y=0$) becomes

\begin{equation}
\frac{d^2N}{dp_Tdy} = \frac{gV}{(2\pi)^2}m_Tp_T\left[1+(q-1)\frac{m_T}{T}\right]^{\frac{-q}{q-1}}.
\end{equation}

In the present study, since we are studying charged hadron spectra in various collision systems, therefore, a transformation from rapidity ($y$) to pseudorapidity ($\eta$) is necessary. This is achieved using the following relation:

\begin{equation}
\frac{dN}{dp_Td\eta} = \frac{dN}{dp_Tdy}\left(\sqrt{1-\frac{m^2}{m_T^2\cosh^2(y)}}\right),
\end{equation}

which simplifies to:

\begin{equation}
\frac{dN}{dp_Td\eta} = \frac{dN}{dp_Tdy}\left(\frac{p_T}{m_T}\right),
\end{equation}

at mid-rapidity.

The charged hadron spectra from various collision systems were fitted with the sum of Tsallis distributions for pions, kaons, and protons. The resulting expression of the fit function is as follows \cite{Azmi2020}:

\begin{equation}
\label{eq:tsa_fitfunc}
\frac{d^2N}{dp_Td\eta} = 2\frac{V}{(2\pi)^2}p_T^2\sum_{i=1}^{3}g_i\left[1+(q-1)\frac{m_{T,i}}{T}\right]^{\frac{-q}{q-1}},
\end{equation}

where $i=\pi^{+},K^{+},p$, and $g_{\pi}=g_{K}=1$ and $g_p=2$ are the degeneracy factors. Factor of $2$ accounts for the antiparticles. This notation is followed throughout the paper.

We have used Eq. (\ref{eq:tsa_fitfunc}) to fit the charged hadron spectra obtained in Pb+Pb collisions at $\sqrt{s_{NN}}$ = 2.76 and 5.02 TeV, Xe+Xe collisions at $\sqrt{s_{NN}}$ = 5.44 TeV, $p$+Pb collisions at $\sqrt{s_{NN}}$ = 5.02 TeV, and $p$+$p$ collisions at $\sqrt{s}$ = 5.02 and 13 TeV, measured by the ALICE Collaboration \cite{2018,Acharya2019,Acharya20192}. The extracted fit parameters are used to calculate various thermodynamic variables. To achieve a fair comparison of these thermodynamic quantities across different collision systems, we have fitted the spectra measured within the same pseudorapidity window of $|\eta| <$  0.8. Furthermore, we have chosen to limit the fits to $p_T < $ 5 GeV/c to investigate bulk properties, as high-$p_T$ particles are generated by hard processes. In a previous study (\cite{Azmi2020}), various thermodynamic quantities were calculated by fitting the spectra in the full $p_T$ range in Pb+Pb collisions at $\sqrt{s_{NN}}$ = 2.76 and 5.02 TeV using Tsallis distribution. It should be noted that the selection of the $p_T$ fitting range impacts the fit parameters and consequently influences the resulting thermodynamic variables \cite{NathPatra2021}.

\begin{figure*}
    \centering
    \includegraphics[width=0.5\textwidth]{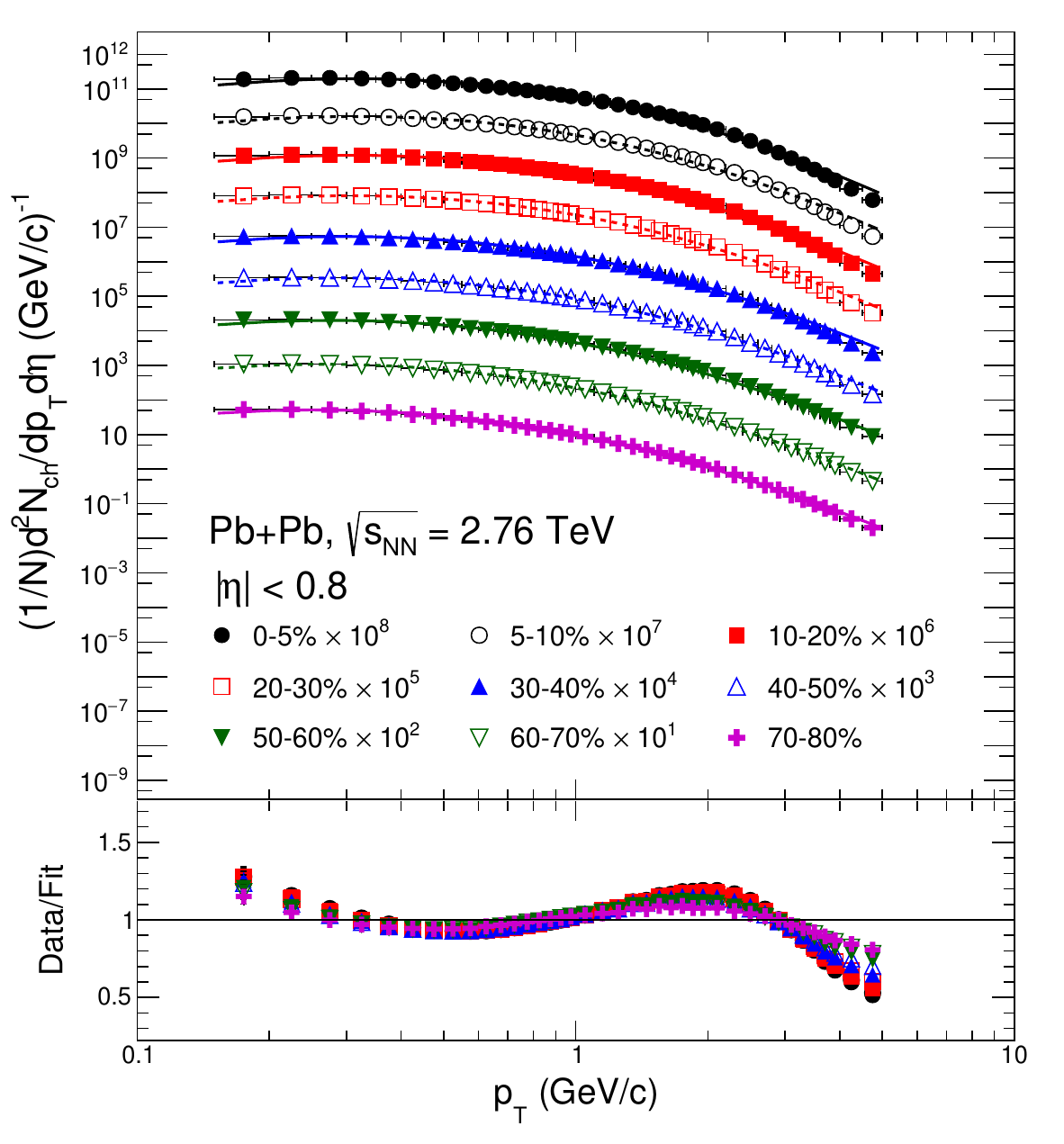}\hfill
    \includegraphics[width=0.5\textwidth]{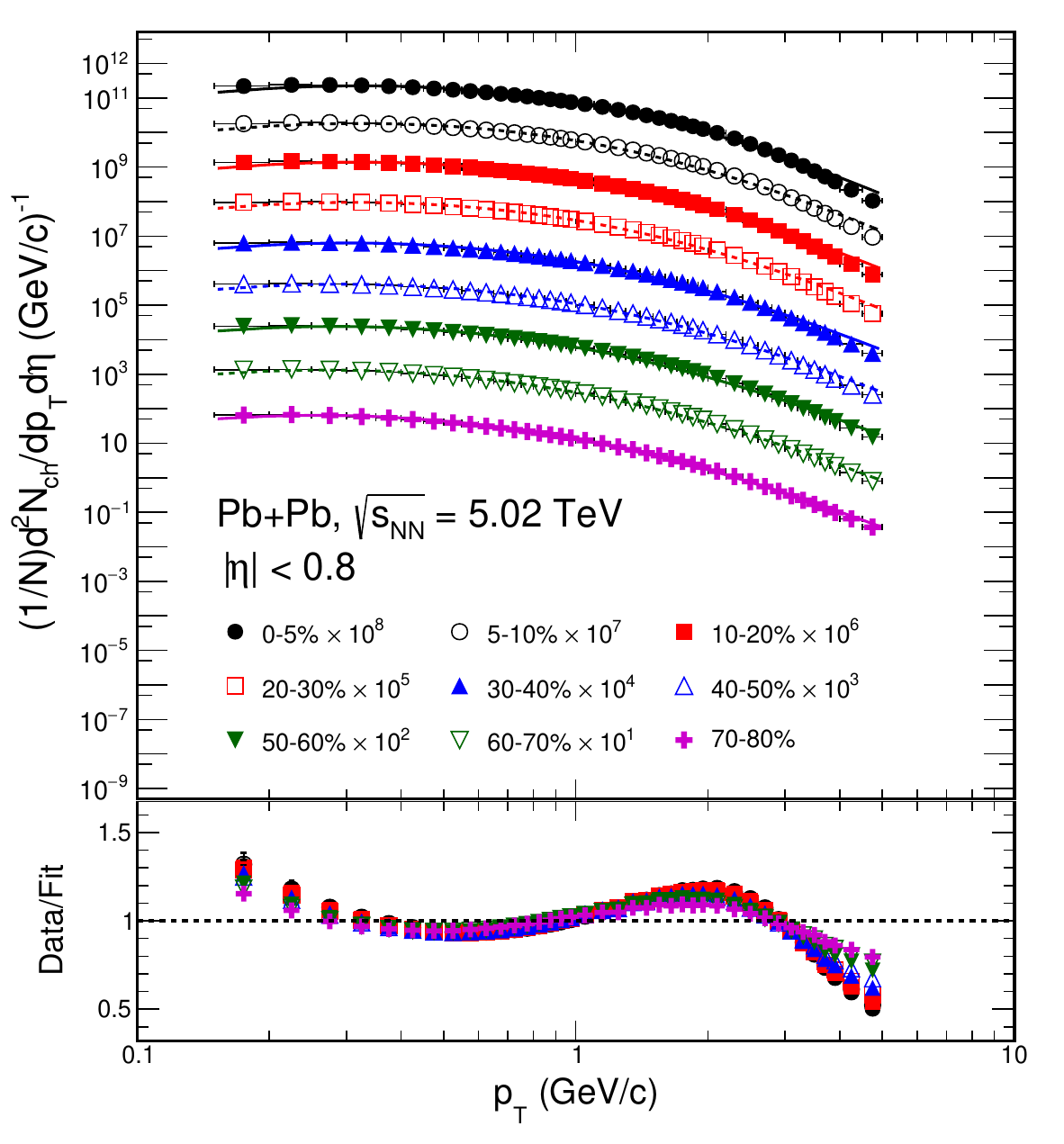}
    \caption{Tsallis fits to the $p_T$-spectra of charged hadrons in different centrality classes of Pb+Pb collisions at $\sqrt{s_{NN}}$ = 2.76 TeV (left) and 5.02 TeV (right). The lower panel of each plot shows the ratio of the data and fit function.}
    \label{fit:tsa_fit_pbpb}
\end{figure*}

Figure \ref{fit:tsa_fit_pbpb} shows the Tsallis fits to the charged hadron spectra in various centrality classes of Pb+Pb collisions at $\sqrt{s_{NN}}$ = 2.76 and 5.02 TeV using Eq. (\ref{eq:tsa_fitfunc}). It is evident that the Tsallis fits provide a more accurate description of the charged hadron spectra in peripheral Pb+Pb collisions compared to central collisions for both collision energies \cite{Azmi2020}. 

\begin{figure*}
    \centering
    \includegraphics[width=0.5\textwidth]{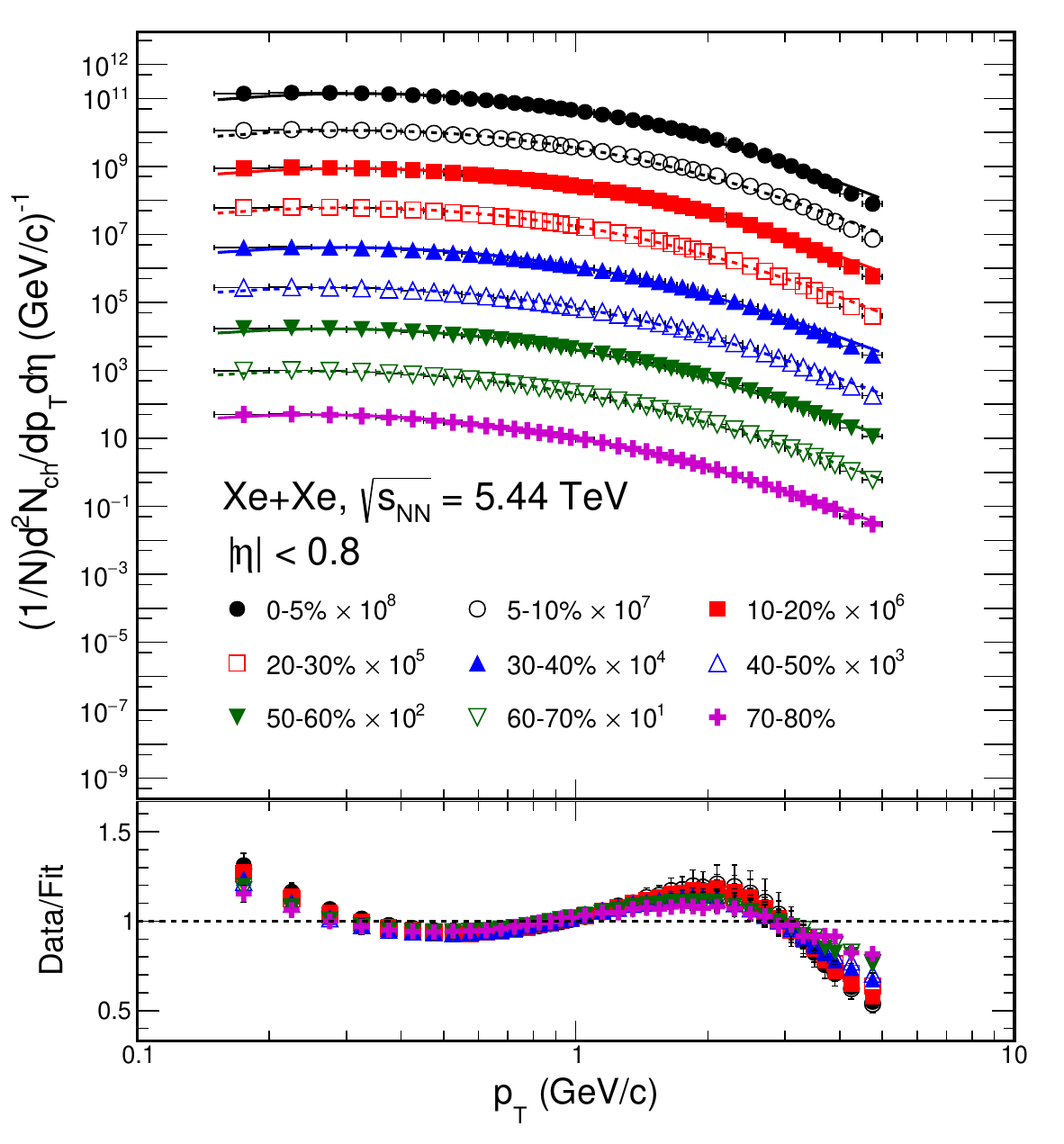}\hfill
    \includegraphics[width=0.5\textwidth]{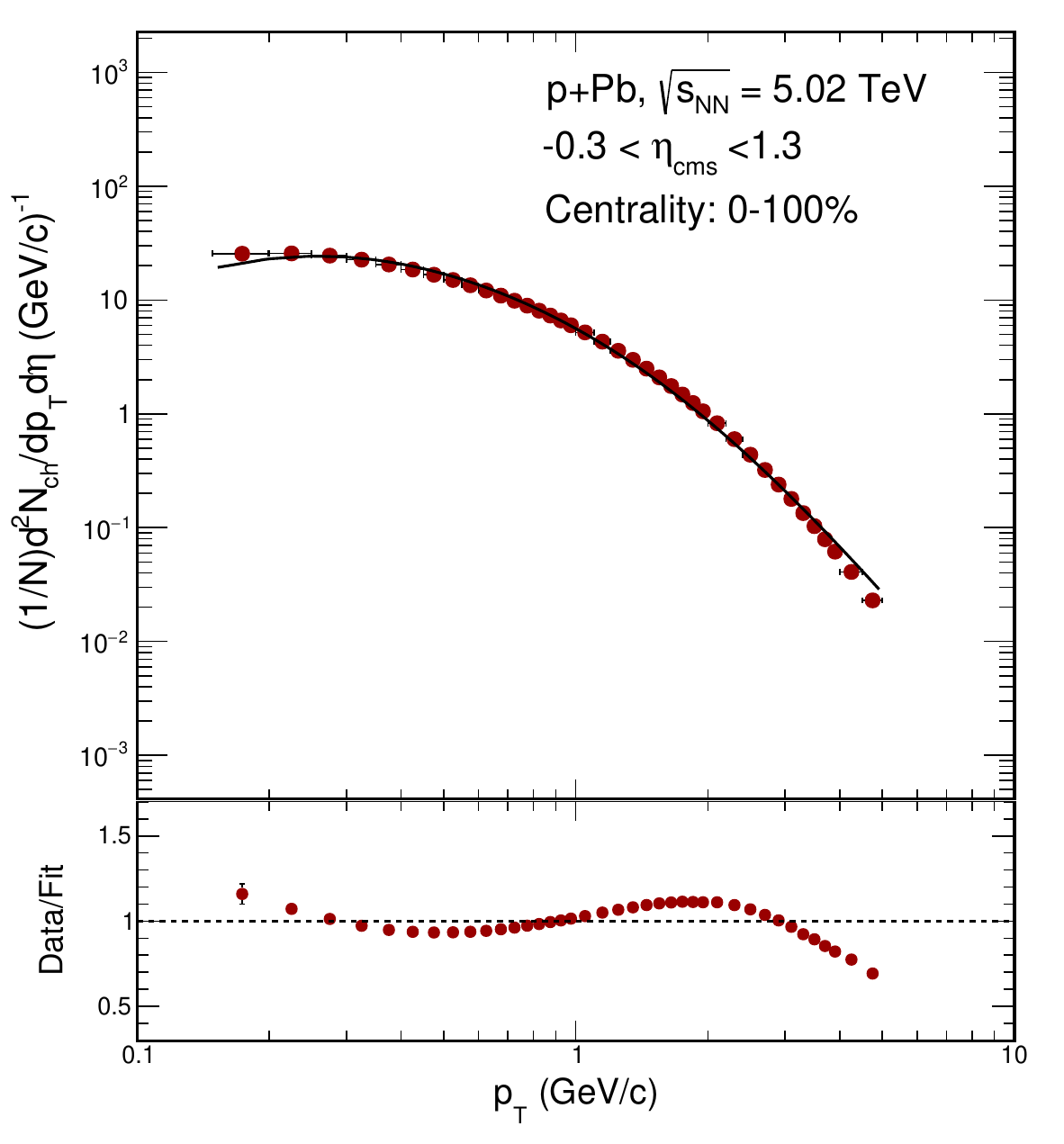}
    \caption{Tsallis fits to the $p_T$-spectra of charged hadrons in different collision centralities of Xe+Xe collisions (left) and 0-100\% centrality of $p$+Pb collisions (right) at $\sqrt{s_{NN}}$ = 5.02 TeV. The lower panel of each plot shows the ratio of the data and fit function.}
    \label{fit:tsa_fit_ppb_xexe}
\end{figure*}

Similary, Fig. \ref{fit:tsa_fit_ppb_xexe} presents the Tsallis distribution fits to the charged hadron spectra in different centrality classes of Xe+Xe collisions at $\sqrt{s_{NN}}$ = 5.44 TeV, and 0-100\% centrality in $p$+Pb collisions at $\sqrt{s_{NN}}$ = 5.02 TeV. As was the case in Pb+Pb collisions, Tsallis fits provide a more accurate description of the charged hadron spectra in peripheral Xe+Xe collisions compared to central collisions. We also observe that the Tsallis distribution provide a good description of the charged particle spectra in $p$+Pb collisions.

\begin{figure*}
    \centering
    \includegraphics[width=0.5\textwidth]{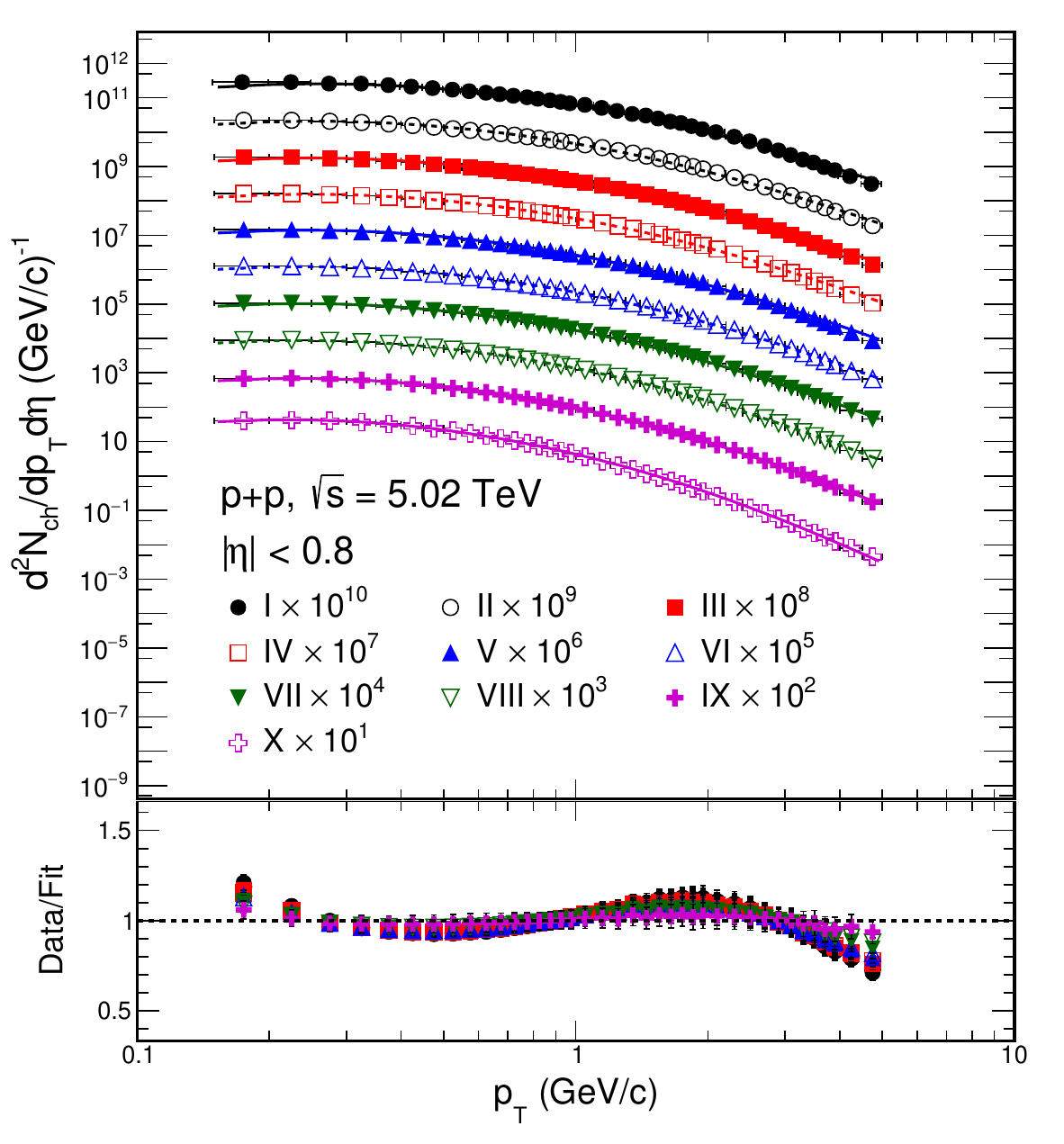}\hfill
    \includegraphics[width=0.5\textwidth]{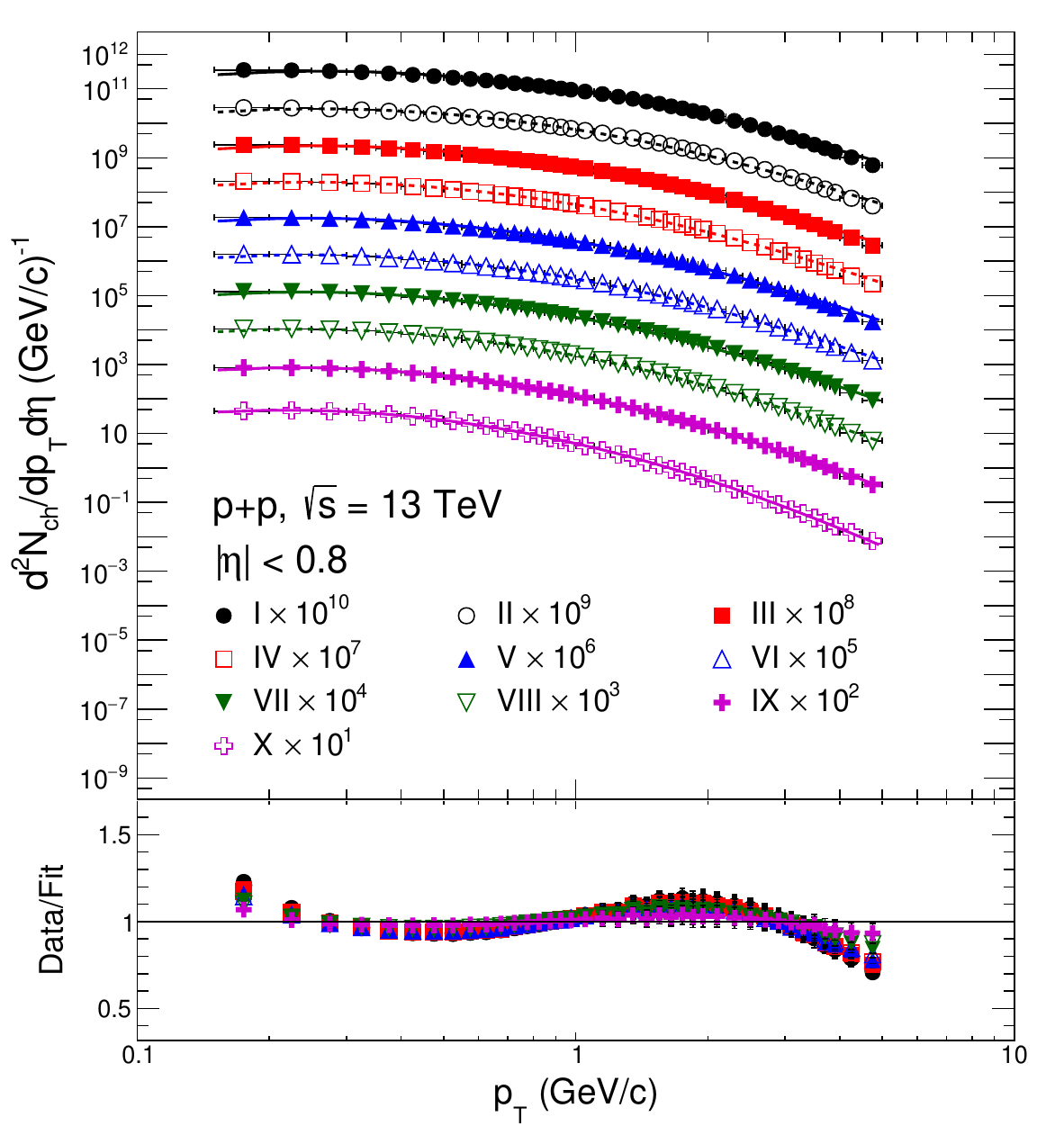}
    \caption{Tsallis fits to the $p_T$-spectra of charged hadrons in different multiplicity classes of $p$+$p$ collisions at $\sqrt{s_{NN}}$ = 5.02 TeV (left) and 13 TeV (right). The lower panel of each plot shows the ratio of the data and fit function.}
    \label{fit:tsa_fit_pp}
\end{figure*}

 Figure \ref{fit:tsa_fit_pp} shows Tsallis fits to the charged hadron spectra measured in different multiplicity classes of $p$+$p$ collisions at $\sqrt{s}$ = 5.02 and 13 TeV. Notably, Tsallis fits provide a better description of the charged hadron spectra in lower multiplicity classes in $p$+$p$ collisions compared to higher multiplicity classes. 

\begin{figure*}
    \centering
    \includegraphics[width=0.5\textwidth]{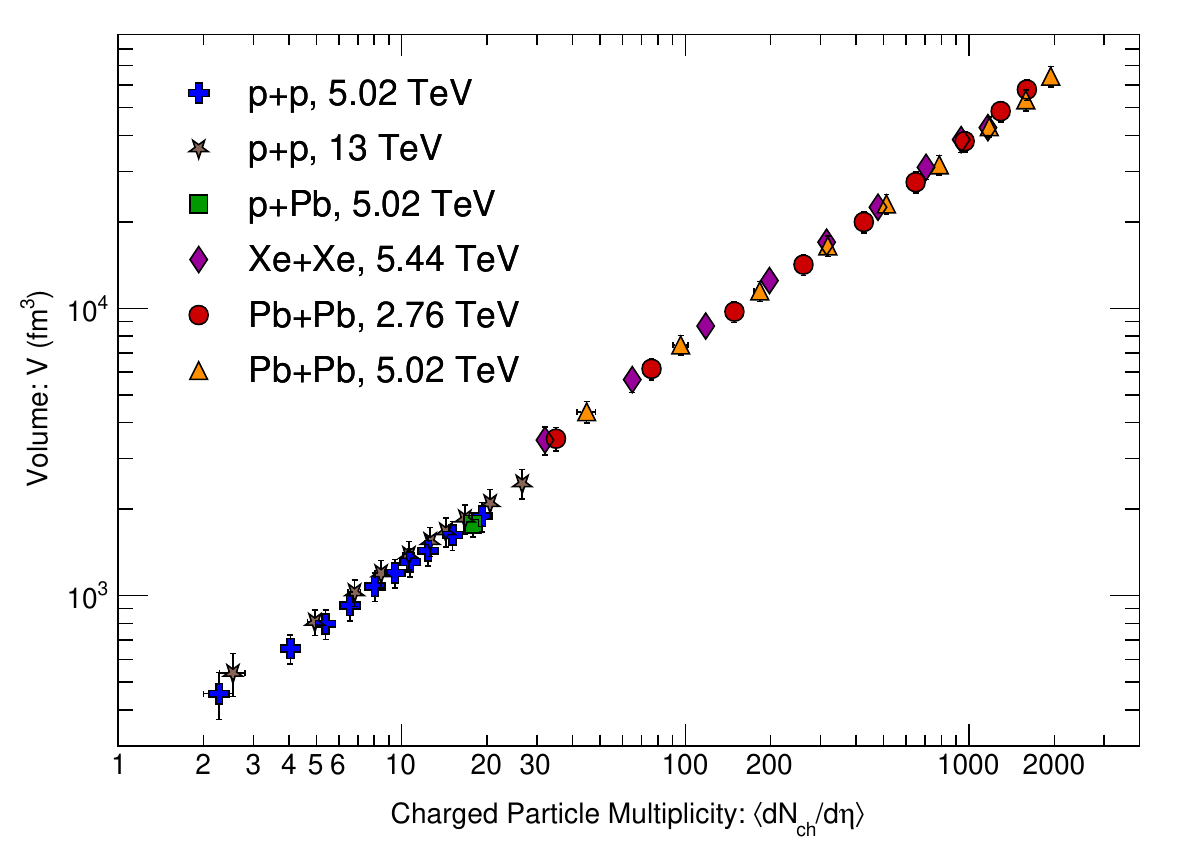}\hfill
    \includegraphics[width=0.5\textwidth]{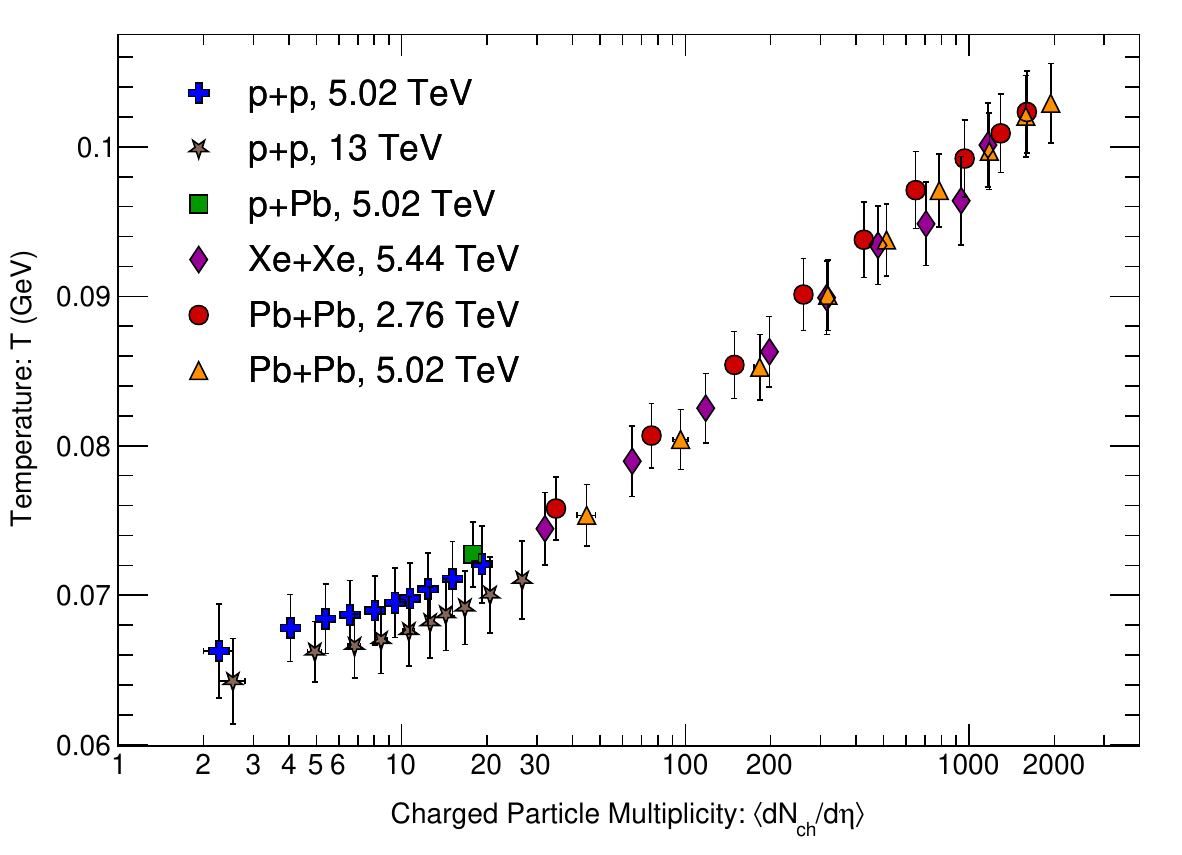}\hfill
    \includegraphics[width=0.5\textwidth]{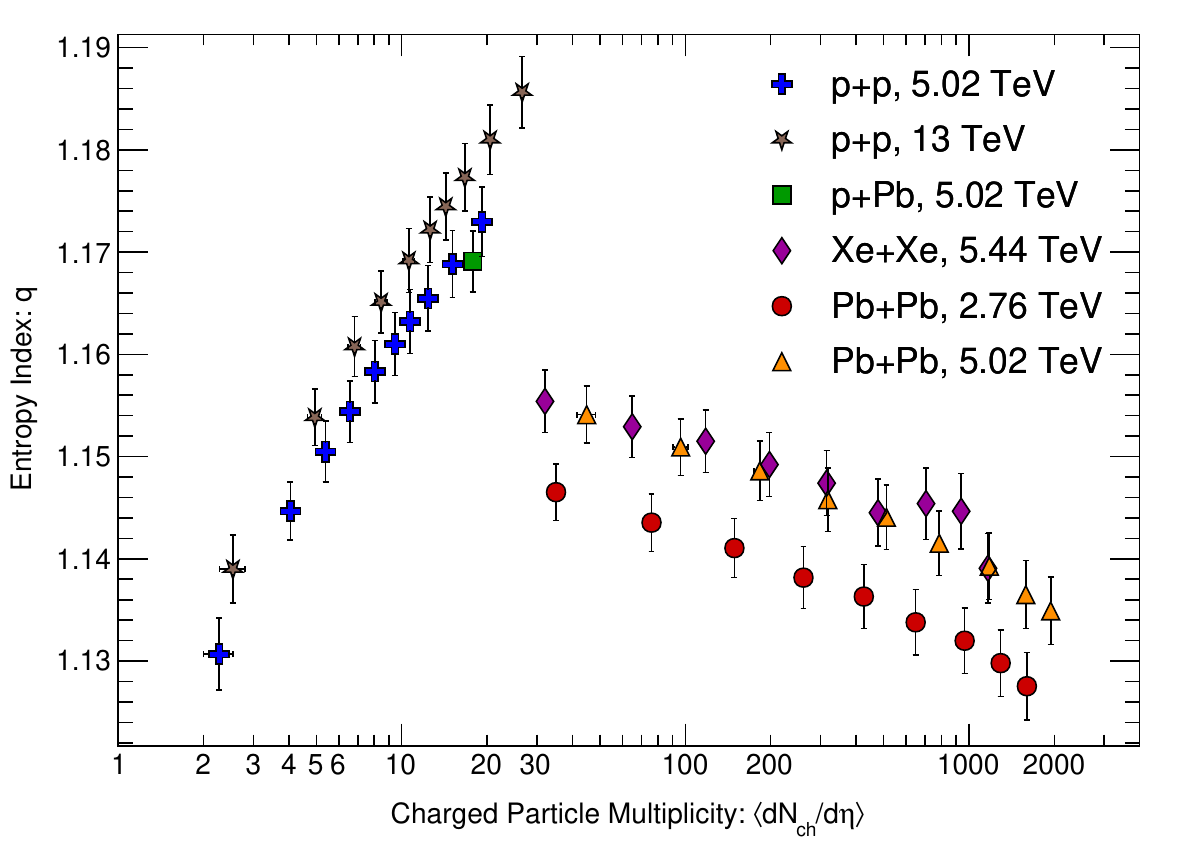}
    \caption{Tsallis parameters (volume, temperature, and $q$) as a function of $\langle dN_{ch}/d\eta \rangle$, obtained by fitting the charged hadron $p_{T}$-spectra in Pb+Pb collisions at $\sqrt{s_{NN}}$ = 2.76 and 5.02 TeV, Xe+Xe collisions at $\sqrt{s_{NN}}$ = 5.44 TeV, $p$+Pb collisions at $\sqrt{s_{NN}}$ = 5.02 TeV, and $p$+$p$ collisions at $\sqrt{s}$ = 5.02 and 13 TeV using Eq. (\ref{eq:tsa_fitfunc}).}
    \label{fit:fit_par}
\end{figure*}

Figure \ref{fit:fit_par} shows the dependence of fit parameters ($V$, $T$, and $q$) on charged particle multiplicity ($\langle dN_{ch}/d\eta \rangle$), which is an indicator of the size of the hadronic system produced in high-energy collisions \cite{Acharya2019,PhysRevC.88.044910,Acharya20192,Acharya2020}. The findings from Fig. \ref{fit:fit_par} lead to the following conclusions:
\begin{itemize}
\item The Tsallis volume, $V$, increases linearly with increasing $\langle dN_{ch}/d\eta \rangle$ and hence is a good indicator of the system size. However, this increase appears to be slower for $p$+$p$ collisions compared to Pb+Pb and Xe+Xe collisions. Similar trend was also observed in femtoscopy studies where small systems have smaller radii compared to large systems extracted with two-pion and three-pion cumulants \cite{GROSSEOETRINGHAUS201422}. This suggests that at the same $\langle dN_{ch}/d\eta \rangle$, the volume of the hadronic system produced in $p$+$p$ collisions is smaller than in Pb+Pb and Xe+Xe collisions. The single measurement in  $p$+Pb collisions falls within the trend observed in $p$+$p$ collisions at the corresponding $\langle dN_{ch}/d\eta \rangle$, suggesting a potential convergence of behavior in small collision systems.

\item The temperature, $T$, increases with increasing $\langle dN_{ch}/d\eta \rangle$. It is observed that $T$ rises sharply in Pb+Pb and Xe+Xe collisions compared to $p$+$p$ collisions with increasing $\langle dN_{ch}/d\eta \rangle$. 

\item The entropy index, $q$, for $p$+$p$ collisions increases with increasing $\langle dN_{ch}/d\eta \rangle$, suggesting a deviation from equilibrium in high-multiplicity $p$+$p$ collisions. The $q$-parameter for $p$+Pb collisions aligns with the trend observed in $p$+$p$ collisions at the corresponding $\langle dN_{ch}/d\eta \rangle$. However, for Pb+Pb and Xe+Xe collisions, $q$-parameter decreases with increasing $\langle dN_{ch}/d\eta \rangle$, suggesting that the system approaches equilibrium more closely in central collisions than in peripheral collisions. It is speculated that the initial energy density induced hot spots caused by the Color Glass Condensate formalism may generate significant temperature fluctuations especially in small systems \cite{Liu2023}. In high multiplicity $p$+$p$ collisions this will cause an increase in the value of $q$, contrary to central heavy-ion collisions. 
We also observe that $q$ increases with increasing center-of-mass energy in both small and large collision systems, indicating a tendency for the system to deviate further from equilibrium in collisions with higher energy.
\end{itemize}

The parameters obtained from the Tsallis fits are used to calculate various thermodynamic quantities, which are explored in the following section. 

\section{\label{sec:thermo}Thermodynamic Variables}
\subsection{\label{sec:energy}Energy}

The Tsallis parameters, $V$, $T$, and $q$ are used to compute the energy density using the following equation:

\begin{equation}
\label{eq:thermo_energy}
\epsilon = 2\sum_{i=1}^{3} g_i \int \frac{d^3p}{(2\pi)^3}E_i\left(1+(q-1)\frac{E_i}{T}\right)^\frac{-q}{q-1},
\end{equation}

\begin{figure*}
    \centering
    \includegraphics[width=0.5\textwidth]{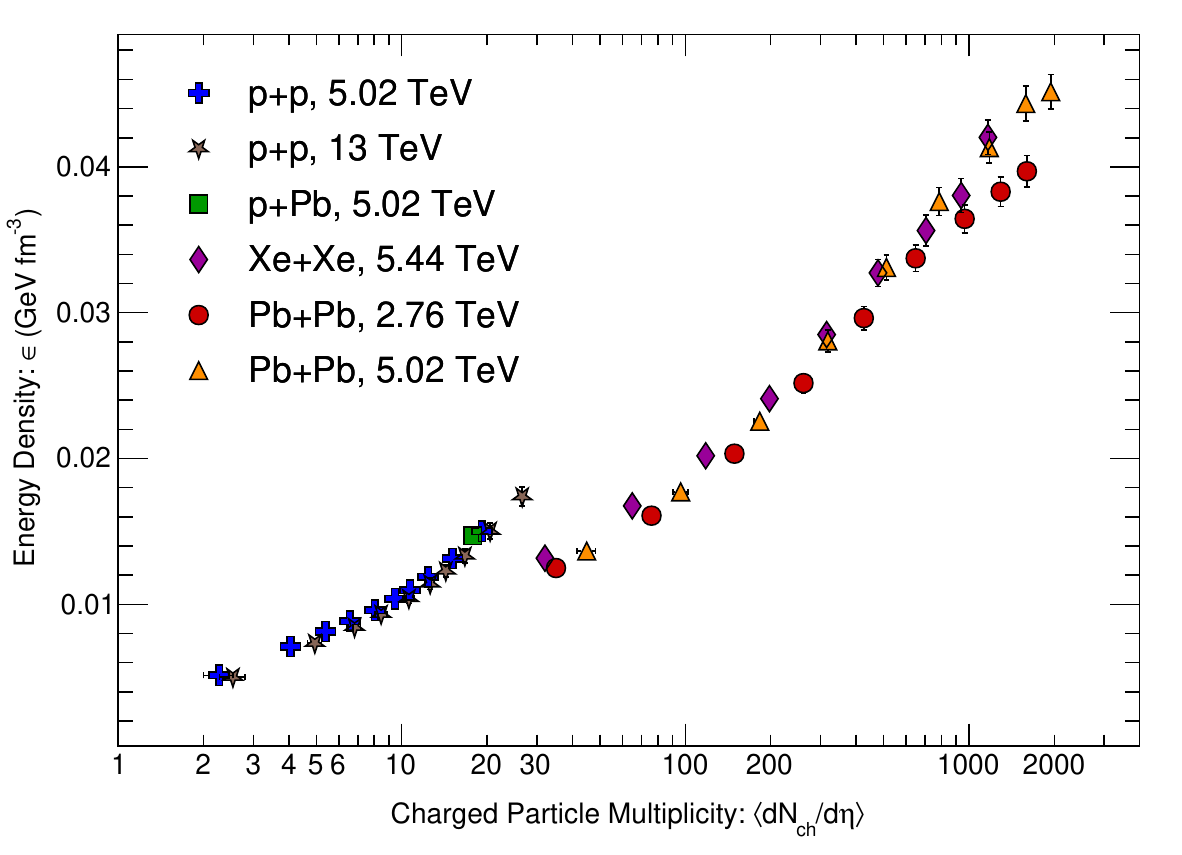}\hfill
    \includegraphics[width=0.5\textwidth]{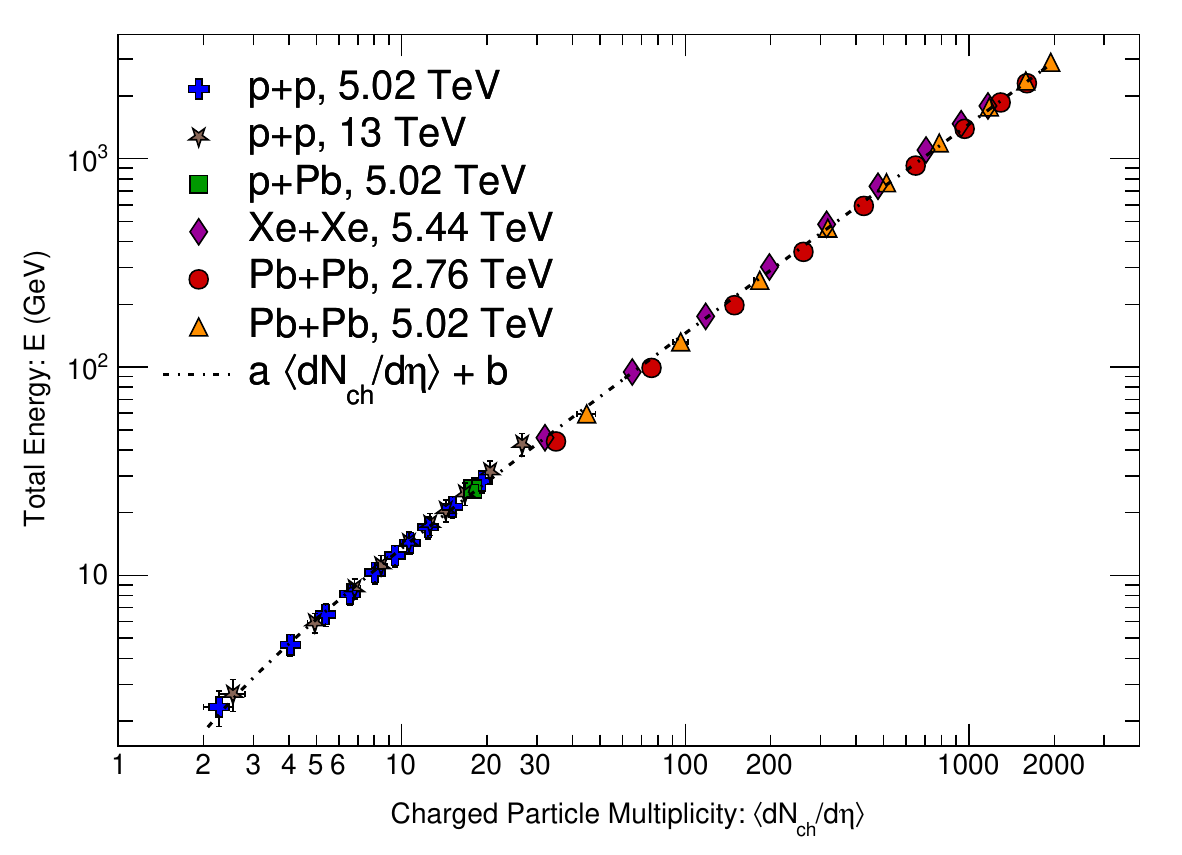}
    \caption{Energy density (left panel) and total energy (right panel) at kinetic freeze-out as a function of $\langle dN_{ch}/d\eta \rangle$ in Pb+Pb collisions at $\sqrt{s_{NN}}$ = 2.76 and 5.02 TeV, Xe+Xe collisions at $\sqrt{s_{NN}}$ = 5.44 TeV, $p$+Pb collisions at $\sqrt{s_{NN}}$ = 5.02 TeV, and $p$+$p$ collisions at $\sqrt{s}$ = 5.02 and 13 TeV calculated using Eq (\ref{eq:thermo_energy}). The expression $a\langle dN_{ch}/d\eta \rangle+b$ is fitted to the total energy and we find $a = 1.463 \pm 0.026$ GeV and $b = -1.163 \pm 0.298$ GeV with $\chi^{2}/ndf$ = 0.207.}
    \label{thermo:energy}
\end{figure*}

The left panel of Fig. \ref{thermo:energy} shows the energy density, $\epsilon$, and the right panel shows the total energy, $E$ (= $\epsilon V$) of the hadronic medium at kinetic freeze-out plotted as a function of $\langle dN_{ch}/d\eta \rangle$ in various collision systems.

It is observed that the energy density and total energy increase with increasing $\langle dN_{ch}/d\eta \rangle$. Furthermore, we find that the energy density exhibits a steeper rise in small collision systems compared to large collision systems, leading to a discontinuity at $\langle dN_{ch}/d\eta \rangle$ $\approx$ 30. This difference might arise because the transverse overlap area in small systems is much smaller than that in large systems for the same event multiplicities \cite{Liu2023,PhysRevC.97.054910}. This is further corroborated by the fact that the discontinuity disappears when $E$ is plotted against $\langle dN_{ch}/d\eta \rangle$ and a linear scaling between the two is observed. This shows that high multiplicity $p$+$p$ collisions are capable of producing energy densities that are comparable to the most peripheral heavy-ion collisions \cite{Sahu20212}.

\subsection{\label{sec:pressure}Pressure}

\begin{figure}
    \centering
    \includegraphics[width=0.5\textwidth]{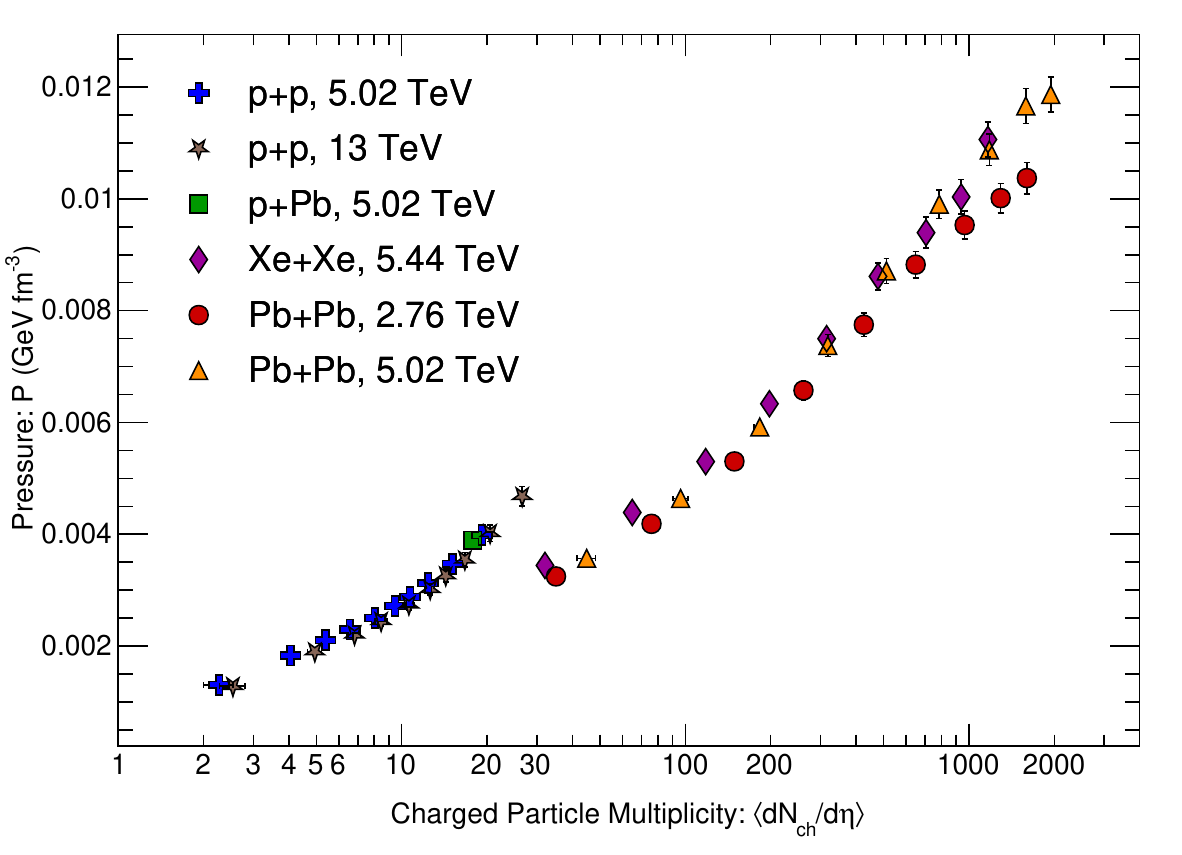}\hfill
    \caption{Pressure at kinetic freeze-out as a function of $\langle dN_{ch}/d\eta \rangle$ in Pb+Pb collisions at $\sqrt{s_{NN}}$ = 2.76 and 5.02 TeV, Xe+Xe collisions at $\sqrt{s_{NN}}$ = 5.44 TeV, $p$+Pb collisions at $\sqrt{s_{NN}}$ = 5.02 TeV, and $p$+$p$ collisions at $\sqrt{s}$ = 5.02 and 13 TeV calculated using Eq (\ref{eq:thermo_pressure}). 
}
    \label{thermo:pressure}
\end{figure}

\begin{figure}
    \centering
    \includegraphics[width=0.5\textwidth]{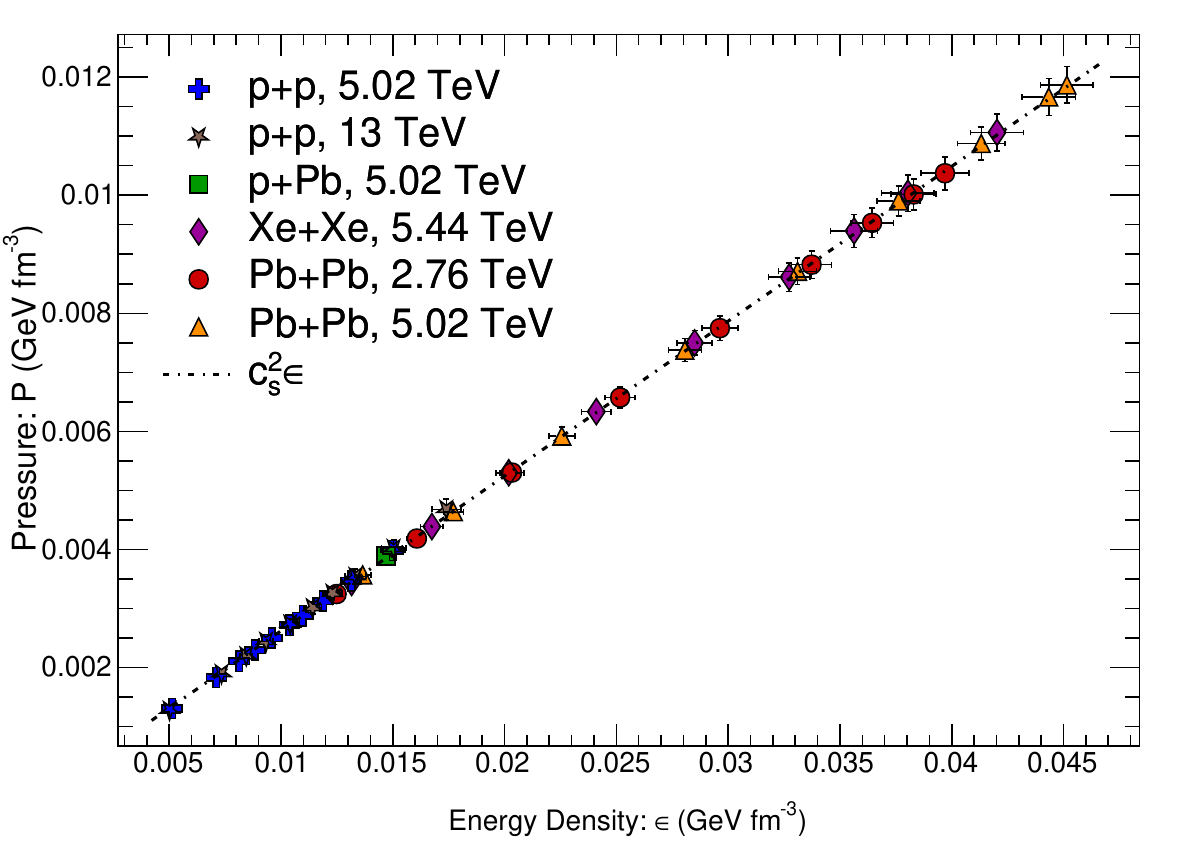}\hfill
    \caption{$P(\epsilon)$ equation of state of charged hadrons in Pb+Pb collisions at $\sqrt{s_{NN}}$ = 2.76 and 5.02 TeV, Xe+Xe collisions at $\sqrt{s_{NN}}$ = 5.44 TeV, $p$+Pb collisions at $\sqrt{s_{NN}}$ = 5.02 TeV, and $p$+$p$ collisions at $\sqrt{s}$ = 5.02 and 13 TeV at kinetic freeze-out. The expression $c_s^2\epsilon$ is fitted to the data and we find $c_s^2 = 0.262 \pm 0.002$ with $\chi^{2}/ndf$ = 0.038.}
    \label{thermo:en_pressure}
\end{figure}

Examining the pressure of hadrnoic medium produced in high energy collisions can provide valuable insights into its equation of state. It can be evaluated using the following equation:

\begin{equation}
\label{eq:thermo_pressure}
P = 2\sum_{i=1}^{3} g_i \int \frac{d^3p}{(2\pi)^3}\frac{p^2}{3E_i}\left(1+(q-1)\frac{E_i}{T}\right)^\frac{-q}{q-1}.
\end{equation}

Figure \ref{thermo:pressure} shows the pressure of the hadronic medium at kinetic freeze-out as a function of $\langle dN_{ch}/d\eta \rangle$, exhibiting an increase with increasing $\langle dN_{ch}/d\eta \rangle$.
It is also observed that the pressure rises more rapidly in $p$+$p$ and $p$+Pb collisions than in Pb+Pb and Xe+Xe collisions, resulting to a discontinuity at $\langle dN_{ch}/d\eta \rangle$ $\approx$ 30. This can possibly be attributed to the higher initial densities in $p$+$p$ and $p$+Pb collisions, which may result in a larger pressure exerted against the surrounding environment \cite{Liu2023,PhysRevC.88.044915,PhysRevC.101.054902}.

Within Landau's hydrodynamic model framework, the equation of state for an ultra-relativistic hadron gas is, $P = c_s^2 \epsilon$, where $c_s^2$ is the square of the speed of sound \cite{PhysRevC.68.064903}. Figure \ref{thermo:en_pressure} shows the linear correlation between pressure and energy density. A linear fit to all the data points results in $c_s^2$ = 0.262 $\pm$ 0.002. A more detailed discussion on $c_s^2$ is presented later.

\subsection{\label{sec:number}Particle number}
\begin{figure*}
    \centering
    \includegraphics[width=0.5\textwidth]{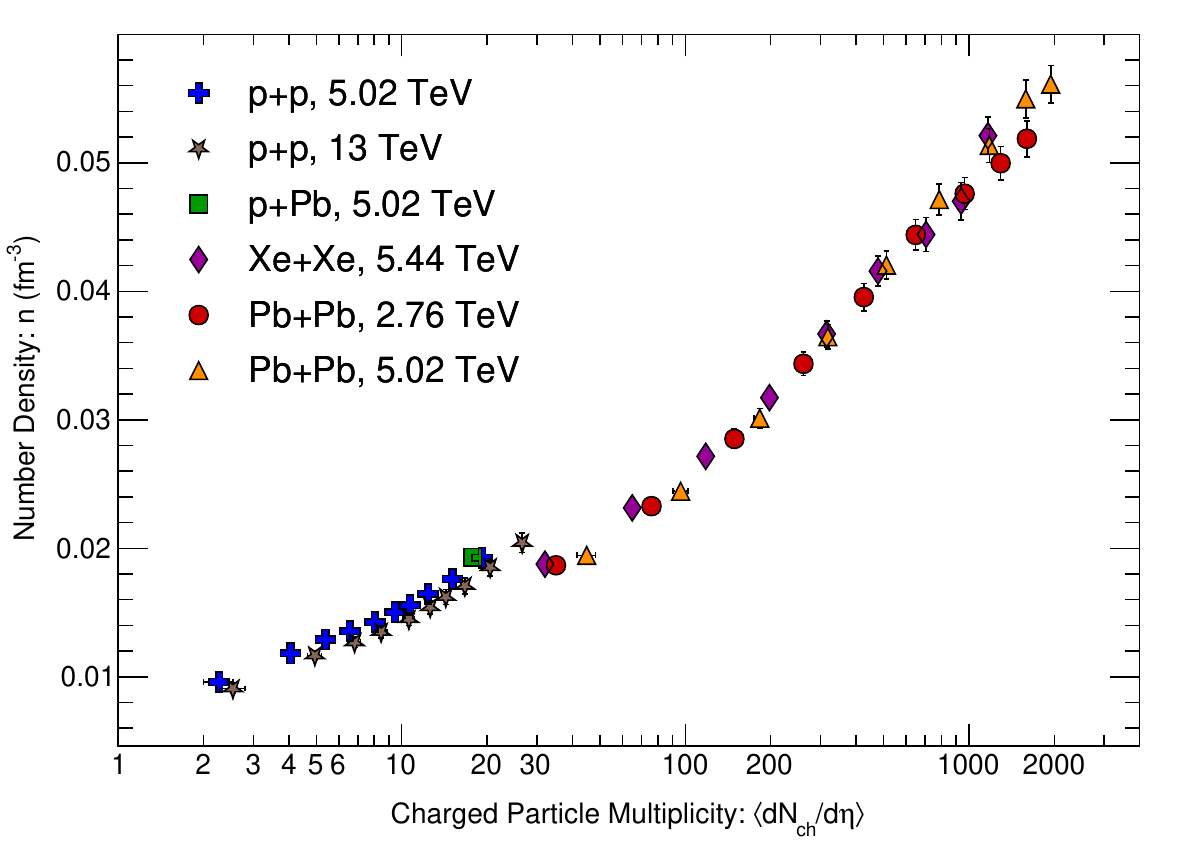}\hfill
    \includegraphics[width=0.5\textwidth]{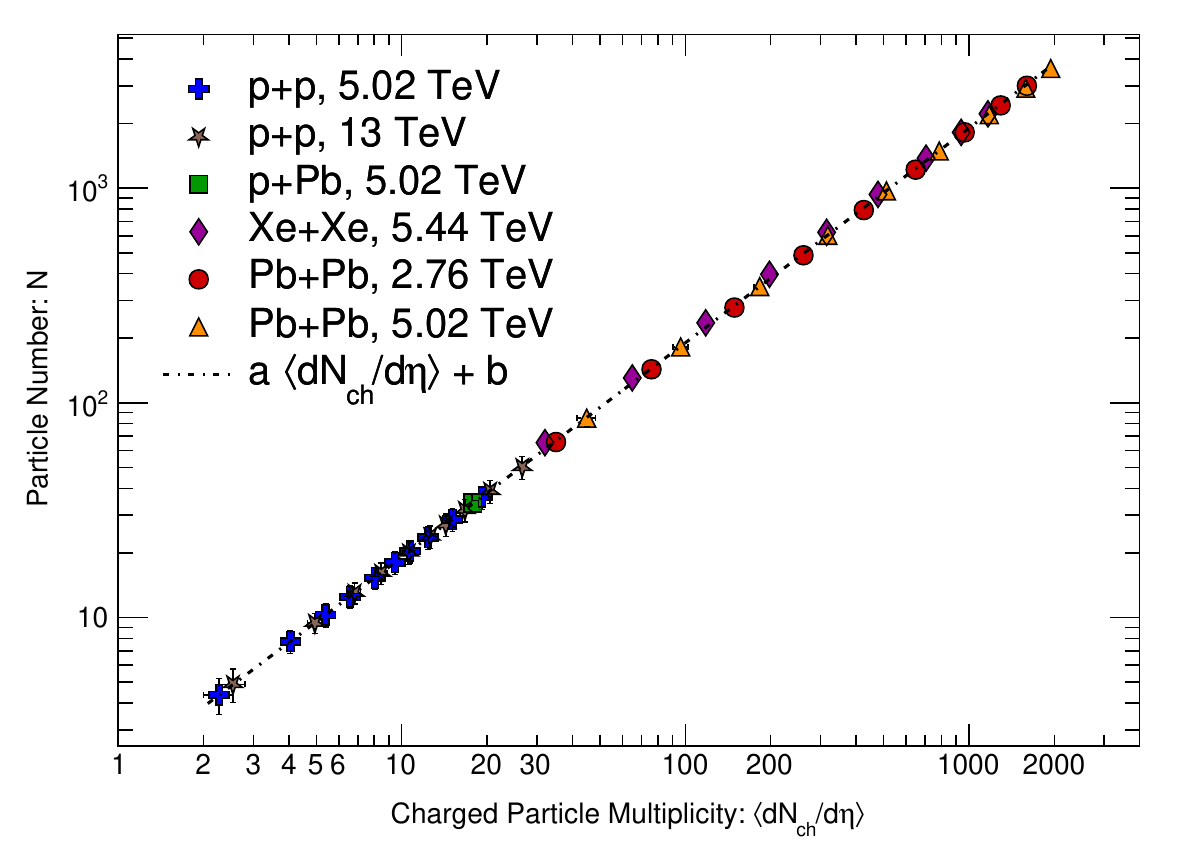}
    \caption{Number density (left panel) and total number of particles (right panel) at kinetic freeze-out as a function of $\langle dN_{ch}/d\eta \rangle$ in Pb+Pb collisions at $\sqrt{s_{NN}}$ = 2.76 and 5.02 TeV, Xe+Xe collisions at $\sqrt{s_{NN}}$ = 5.44 TeV, $p$+Pb collisions at $\sqrt{s_{NN}}$ = 5.02 TeV, and $p$+$p$ collisions at $\sqrt{s}$ = 5.02 and 13 TeV calculated using Eq (\ref{eq:number}). The expression $a\langle dN_{ch}/d\eta \rangle$ is fitted to the total number of particles and we find $a = 1.904	\pm 0.030$ and $b = 0.035	\pm 0.448$ with $\chi^{2}/ndf$ = 0.043.}
    \label{thermo:number}
\end{figure*}

The particle number density can be evaluated from the expression:
 \begin{equation}
 \label{eq:number}
    n = 2\sum_{i=1}^{3} g_i \int \frac{d^3p}{(2\pi)^3}\left(1+(q-1)\frac{E_i}{T}\right)^\frac{-q}{q-1},
\end{equation}   
Figure \ref{thermo:number} shows increase in the number density, $n$, and total number of particles, $N$ (= $nV$), with increasing $\langle dN_{ch}/d\eta \rangle$. It is observed that the increase in particle density is more pronounced for $p$+$p$ and $p$+Pb collisions than for Pb+Pb and Xe+Xe collisions, which could be attributed to the smaller volume of the former collision systems. Moreover, a linear increase of $N$ with $\langle dN_{ch}/d\eta \rangle$ further strengthens the argument that particle number density is closely related to the volume of the system.

\subsection{\label{sec:entropy}Entropy}
The entropy plays a major role in understanding the evolution of the system produced in heavy-ion collisions. Entropy density can be evaluated using the following expression:
\begin{eqnarray}
    \label{eq:thermo_entropy}
    s = 2\sum_{i=1}^{3} g_i \int \frac{d^3p}{(2\pi)^3}\left[\frac{E_i}{T}\left(1+(q-1)\frac{E_i}{T}\right)^\frac{-q}{q-1} \right.\nonumber\\ \left. + \left(1+(q-1)\frac{E_i}{T}\right)^\frac{-1}{q-1}\right].
\end{eqnarray}
\begin{figure*}
    \centering
    \includegraphics[width=0.5\textwidth]{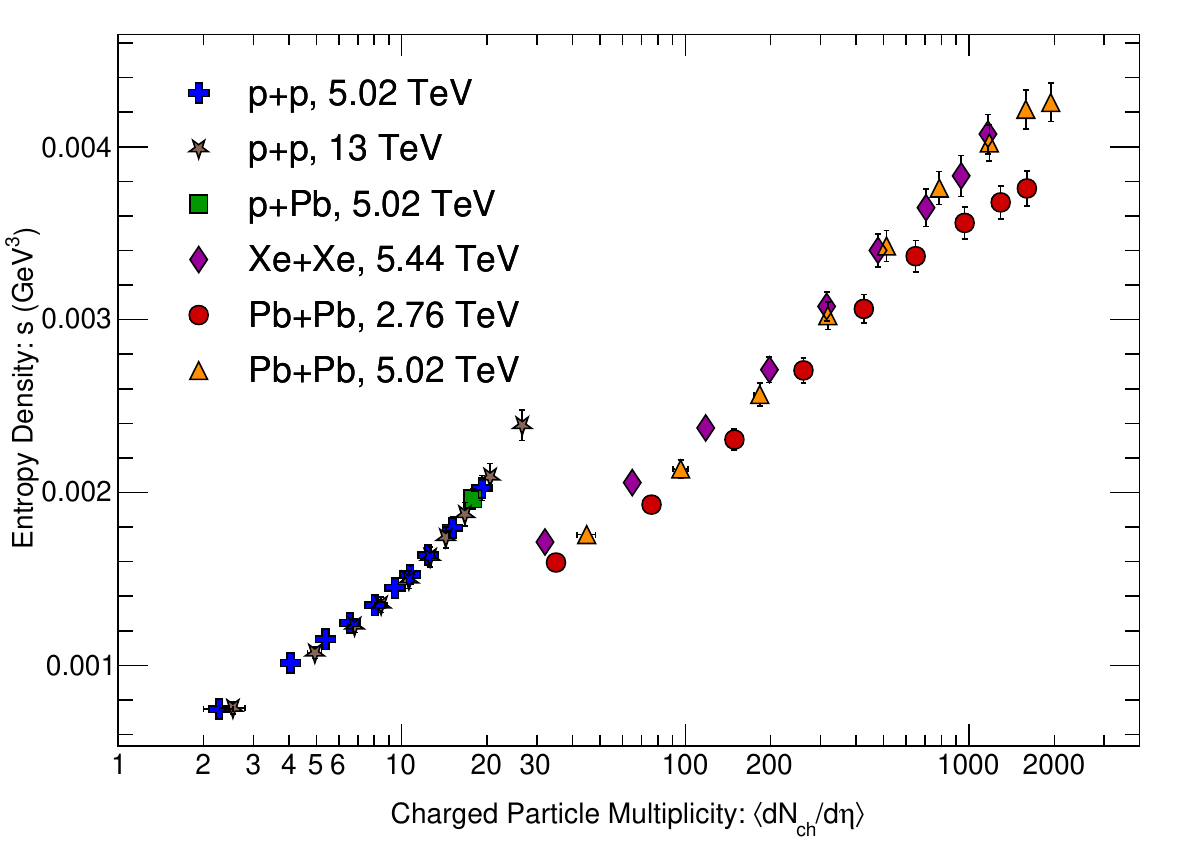}\hfill
    \includegraphics[width=0.5\textwidth]{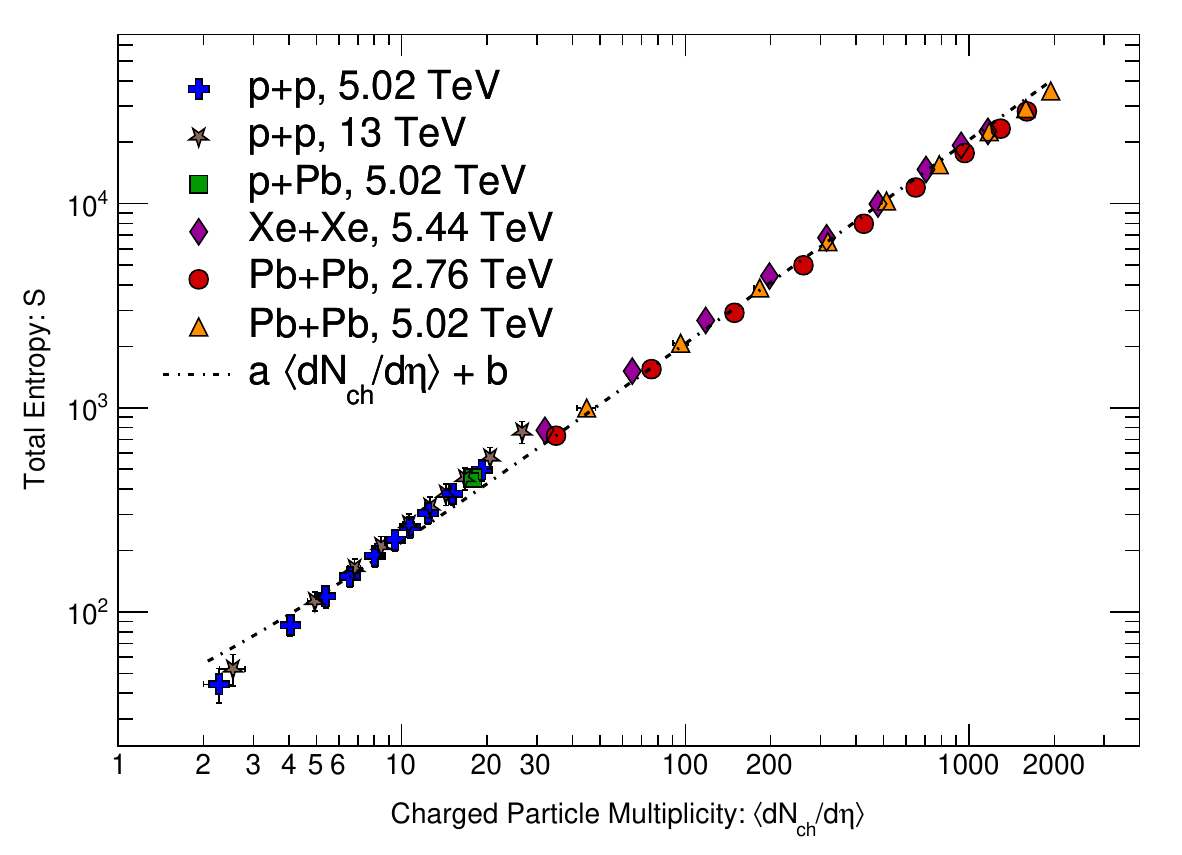}
    \caption{Entropy density (left panel) and total entropy (right panel) at kinetic freeze-out as a function of $\langle dN_{ch}/d\eta \rangle$ in Pb+Pb collisions at $\sqrt{s_{NN}}$ = 2.76 and 5.02 TeV, Xe+Xe collisions at $\sqrt{s_{NN}}$ = 5.44 TeV, $p$+Pb collisions at $\sqrt{s_{NN}}$ = 5.02 TeV, and $p$+$p$ collisions at $\sqrt{s}$ = 5.02 and 13 TeV calculated using Eq (\ref{eq:thermo_entropy}). The expression $a\langle dN_{ch}/d\eta \rangle+b$ is fitted to total entropy and we find $a = 20.489 \pm 0.368$ and $b = 15.012 \pm 4.939$ with $\chi^{2}/ndf$ = 1.702. }
    \label{thermo:entropy}
\end{figure*}
In Fig. \ref{thermo:entropy}, we observe that the entropy density, $s$, and total entropy, $S$ (= $sV$), increase with increasing $\langle dN_{ch}/d\eta \rangle$. This can be attributed to an increase in both the particle number and the volume of the system. Additionally, we note that the entropy density increases more rapidly in $p$+$p$ and $p$+Pb collisions than in Pb+Pb and Xe+Xe collisions, leading to a discontinuity at $\langle dN_{ch}/d\eta \rangle$ $\approx$ 30. It can be inferred from an almost linear relationship between $S$ and $\langle dN_{ch}/d\eta \rangle$ that the dissimilarity in the rate of increase may be ascribed to the volume of the hadronic system. We have also explicitly verified that the thermodynamic relation, $\epsilon + P = Ts$, holds. 

\subsection{\label{sec:path_knudsen}Mean free path and Knudsen number}

\begin{figure*}
    \centering
    \includegraphics[width=0.5\textwidth]{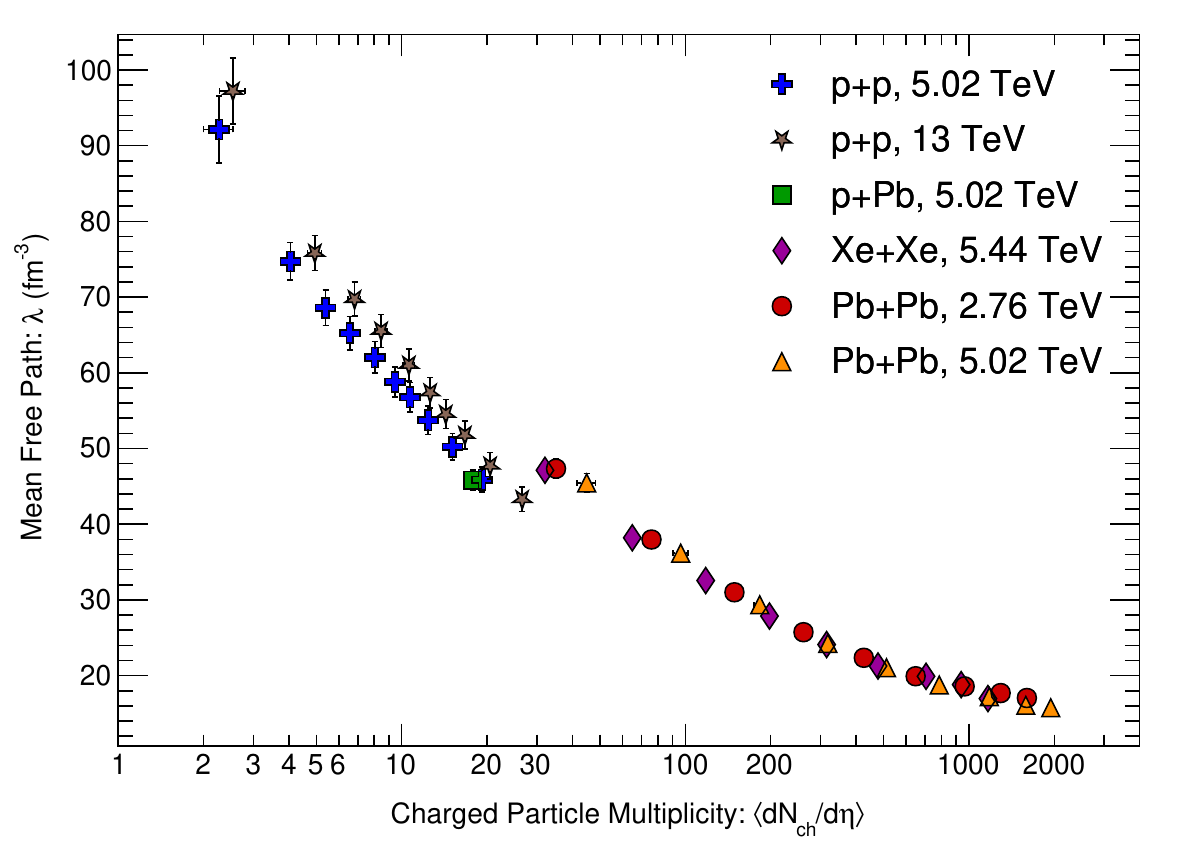}\hfill
    \includegraphics[width=0.5\textwidth]{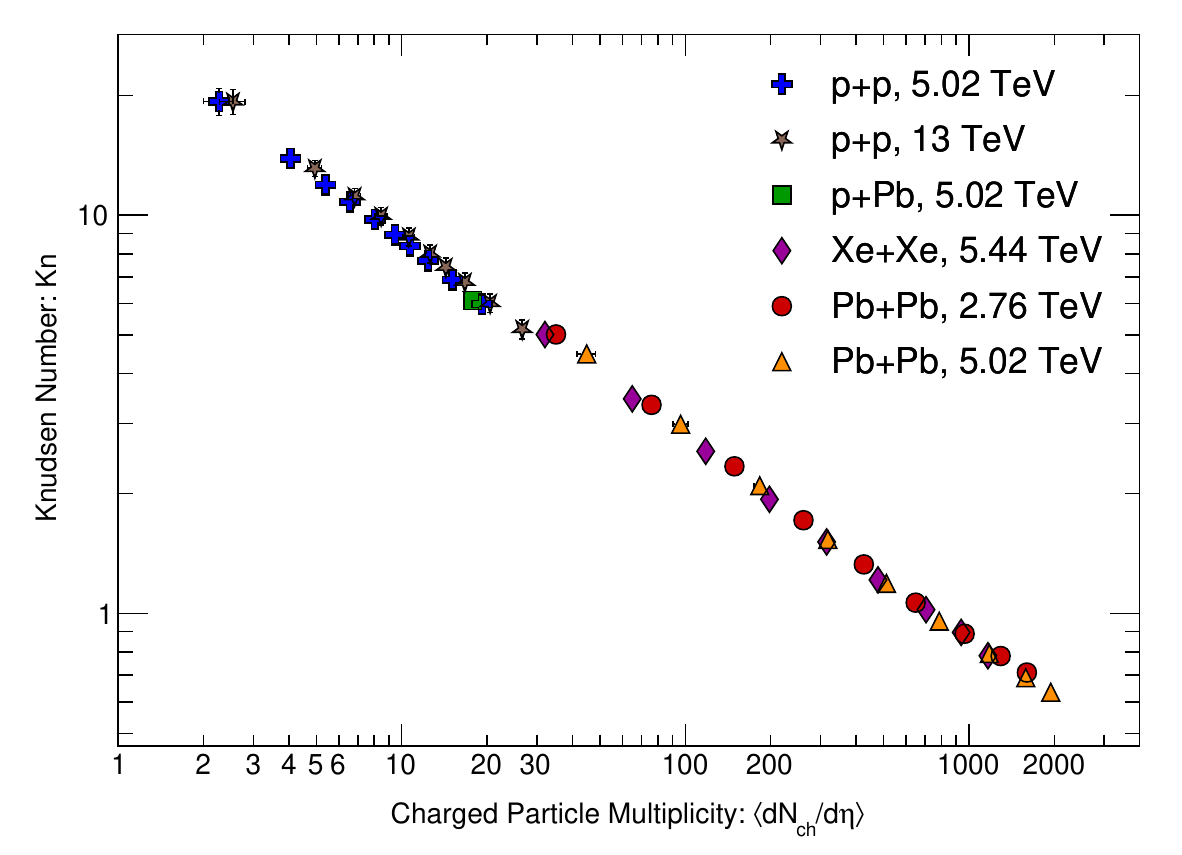}
    \caption{Mean free path (left panel) and Knudsen number (right panel) at kinetic freeze-out as a function of $\langle dN_{ch}/d\eta \rangle$ in Pb+Pb collisions at $\sqrt{s_{NN}}$ = 2.76 and 5.02 TeV, Xe+Xe collisions at $\sqrt{s_{NN}}$ = 5.44 TeV, $p$+Pb collisions at $\sqrt{s_{NN}}$ = 5.02 TeV, and $p$+$p$ collisions at $\sqrt{s}$ = 5.02 and 13 TeV. Mean free path is evaluated using Eq (\ref{eq:lambda}) and the definition of Knudsen number is given in the text.}
    \label{thermo:path_lambda}
\end{figure*}

The mean free path ($\lambda$) of a system is defined as the average distance that a particle travels between two successive collisions in the system. Mathematically, the mean free path is expressed as \cite{Sahu2021}: 
\begin{equation}
\label{eq:lambda}
\lambda = \frac{1}{n\sigma},
\end{equation}
where $n$ is the particle density, and $\sigma$ represents the scattering cross-section. In the specific case of hard core hadron radius approximation, with $r_h=0.3$ fm, $\sigma = 4\pi r_h^2$ \cite{Sahu2021,PhysRevC.92.035203,Bugaev2013}.
Furthermore, Knudsen number ($Kn$) is a dimensionless parameter, defined as the ratio of the mean free path ($\lambda$) to the radius, $R$ (= $(3V/4\pi)^{1/3}$), of the hadronic system \cite{PhysRevC.96.044901,https://doi.org/10.48550/arxiv.1905.06532}. A small value of $Kn$ signifies that the gas system is in the continuum region, allowing the application of hydrodynamics. Generally, when $Kn<0.5$, the fluid-dynamical criterion is satisfied, indicating fluid-like behavior, whereas a value greater than unity indicates that the particles are mostly free-streaming \cite{PhysRevC.79.014906,PhysRevC.82.024910,PhysRevC.79.054914}.

Figure \ref{thermo:path_lambda}, shows both $\lambda$ and $Kn$ decrease with increasing $\langle dN_{ch}/d\eta \rangle$. The decrease in $\lambda$ results from the increase in the particle number density with larger system sizes. Consequently, the hadrons experience more frequent collisions, leading to a decrease in the average distance between successive collisions. The decrease appears to be more rapid in $p$+$p$ collisions because of the smaller volume of the hadronic system compared to large collision systems at the same charged particle multiplicity.

It was observed that, except for collisions with large $\langle dN_{ch}/d\eta \rangle$, the Knudsen number ($Kn$) for the hadronic gas at KFO was greater than 1. This suggests that the mean free path ($\lambda$) is comparable to the system size ($R$), indicating that the hadronic system at KFO is not in the continuum phase. Instead, the particles are predominantly free-streaming.

\subsection{\label{sec:heat}Heat capacity}
\begin{figure}
    \centering
    \includegraphics[width=0.5\textwidth]{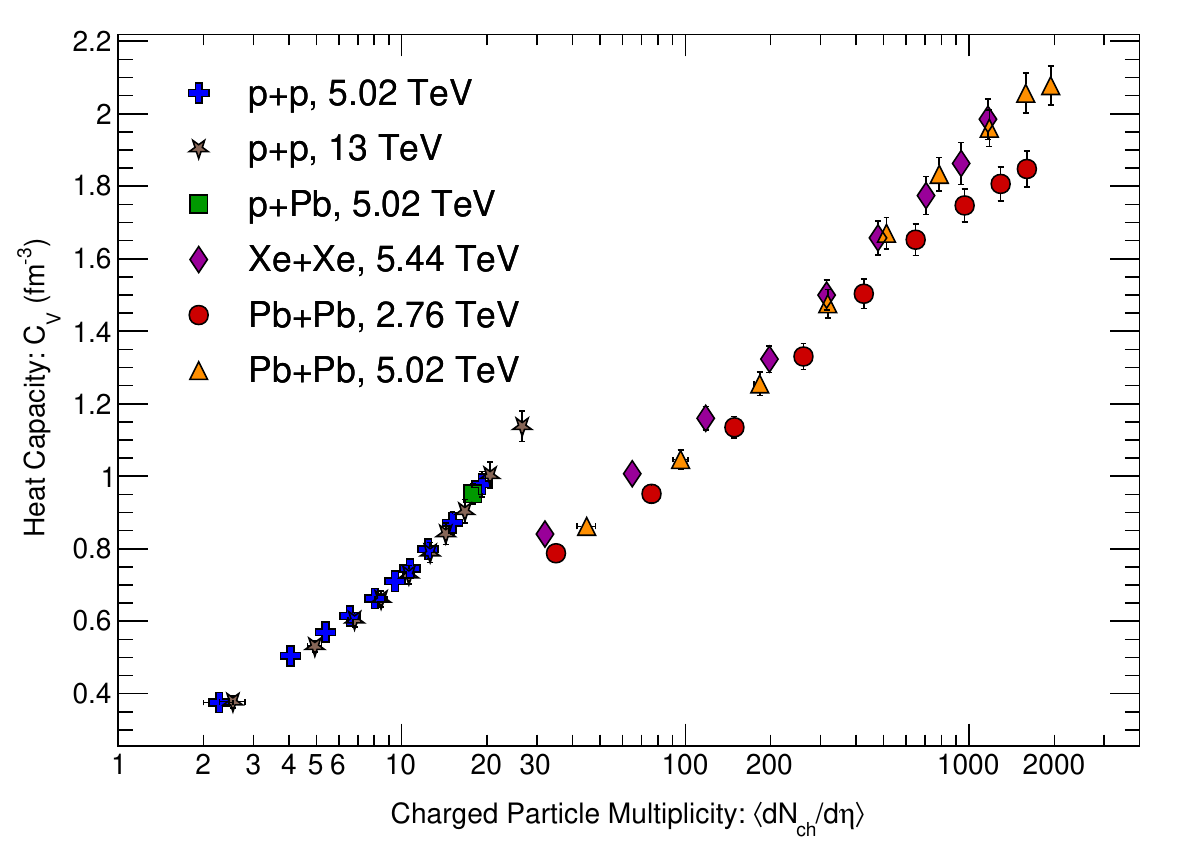}\hfill
    \caption{Heat capacity at kinetic freeze-out as a function of $\langle dN_{ch}/d\eta \rangle$ in Pb+Pb collisions at $\sqrt{s_{NN}}$ = 2.76 and 5.02 TeV, Xe+Xe collisions at $\sqrt{s_{NN}}$ = 5.44 TeV, $p$+Pb collisions at $\sqrt{s_{NN}}$ = 5.02 TeV, and $p$+$p$ collisions at $\sqrt{s}$ = 5.02 and 13 TeV calculated using Eq (\ref{eq:thermo_heat}). 
}
    \label{thermo:heat}
\end{figure}

The heat capacity ($C_V$) of a system quantifies the amount of energy required to increase its temperature by one unit. Mathematically, this can be expressed as \cite{Sahu2021,Khuntia2016}:
\begin{equation}
\label{eq:thermo_heat}
C_V = \left.\frac{\partial \epsilon}{\partial T}\right\vert_V
\end{equation}
Figure \ref{thermo:heat} shows the variation in $C_V$ as a function of $\langle dN_{ch}/d\eta \rangle$. We observe an increase in $C_V$ with increasing $\langle dN_{ch}/d\eta \rangle$. This could be a consequence of the increase in the particle number with $\langle dN_{ch}/d\eta \rangle$. As the number of particles in a system increases, a higher amount of energy is required to increase its temperature by one unit. 

We also observe a steeper increase in $C_V$ with $\langle dN_{ch}/d\eta \rangle$ in the case of $p$+$p$ collisions, leading to a discontinuity at $\langle dN_{ch}/d\eta \rangle$ $\approx$ 30. This might be due to the higher value of the $q$-parameter in $p$+$p$ collisions. Further the system is away from equilibrium, as indicated by a higher $q$-value, more is the change in its internal energy as temperature rises, resulting in a higher $C_V$ \cite{Sahu2021}. 

\subsection{\label{sec:compress}Isothermal compressiblity}
\begin{figure}
    \centering
    \includegraphics[width=0.5\textwidth]{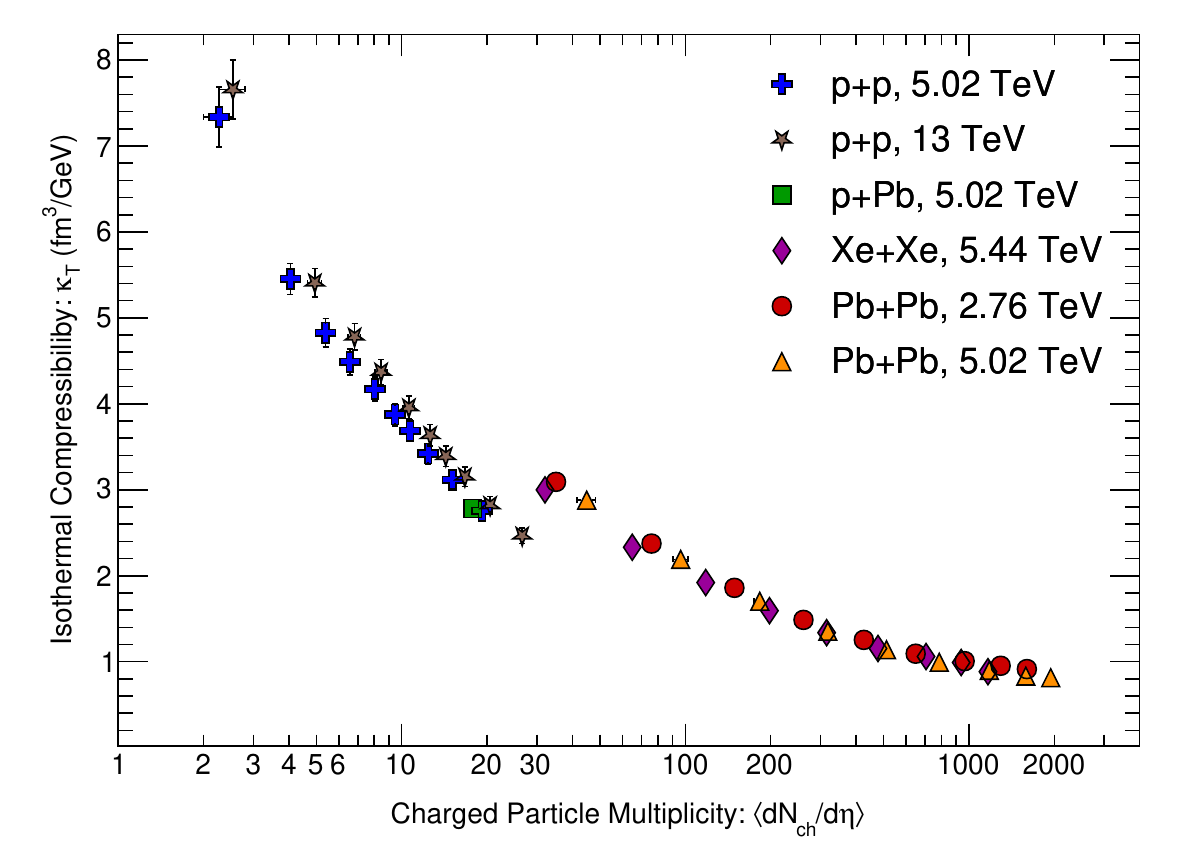}\hfill
    \caption{Isothermal compressibility at kinetic freeze-out as a function of $\langle dN_{ch}/d\eta \rangle$ in Pb+Pb collisions at $\sqrt{s_{NN}}$ = 2.76 and 5.02 TeV, Xe+Xe collisions at $\sqrt{s_{NN}}$ = 5.44 TeV, $p$+Pb collisions at $\sqrt{s_{NN}}$ = 5.02 TeV, and $p$+$p$ collisions at $\sqrt{s}$ = 5.02 and 13 TeV calculated using Eq. (\ref{eq:compress}). 
}
    \label{thermo:compress}
\end{figure}
The isothermal compressibility ($\kappa_T$) of a system is a measure of the extent to which its volume changes in response to an external pressure. For an ideal fluid that is both incompressible and non-viscous, $\kappa_T$ is $0$. It can be evaluated from the expression \cite{Sahu2021,Jain2023,PhysRevC.100.014910}:
 \begin{equation}
 \label{eq:compress}
\kappa_T = \frac{\partial n/\partial \mu}{n^2}.
\end{equation}  
Figure \ref{thermo:compress} shows isothermal compressibility as a function of $\langle dN_{ch}/d\eta \rangle$.  We observe that $\kappa_T$ decreases with increasing $\langle dN_{ch}/d\eta \rangle$, suggesting that the system approaches near ideal behavior with increasing multiplicity. This can be explained by the fact that higher multiplicity classes consist of a greater number of particles, thus requiring higher pressure to achieve a small change in volume \cite{Jain2023}.

\subsection{\label{sec:expansion}Expansion coefficient}
\begin{figure}
    \centering
    \includegraphics[width=0.5\textwidth]{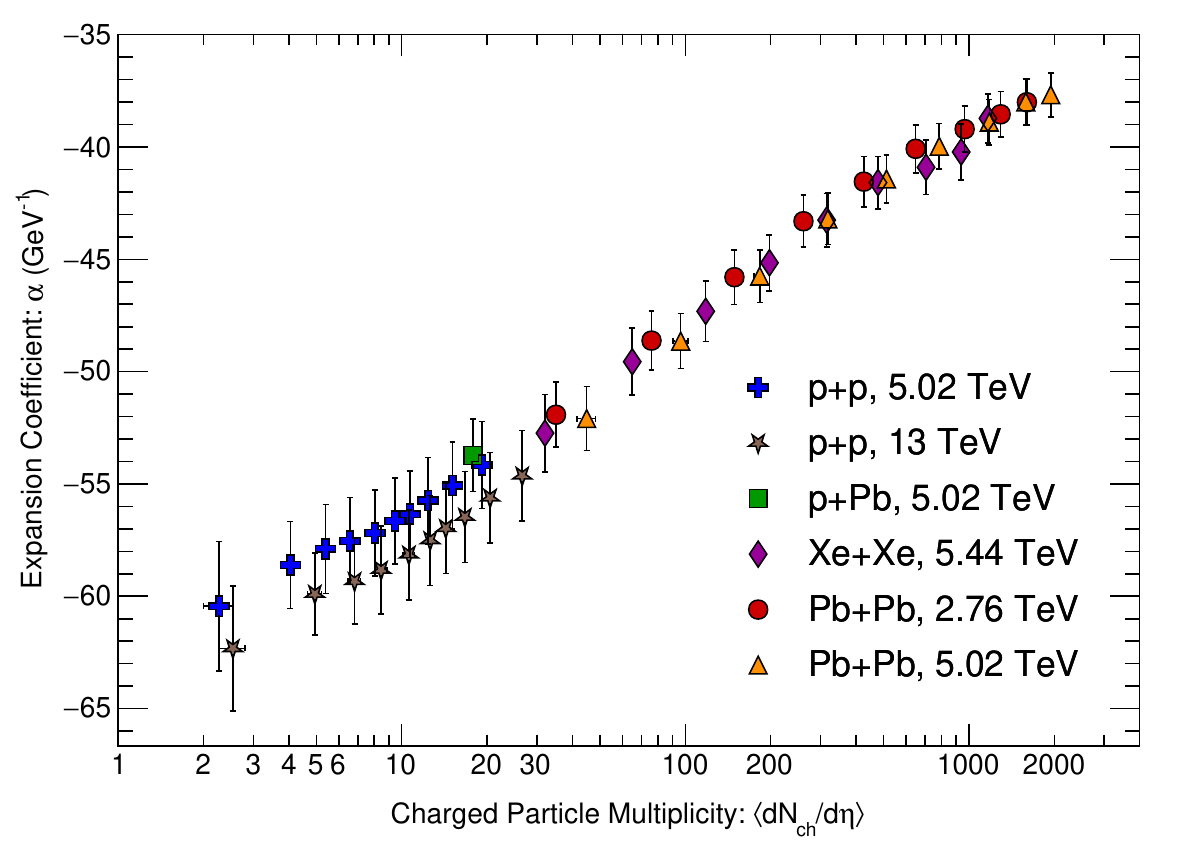}\hfill
    \caption{Expansion coefficient at kinetic freeze-out as a function of $\langle dN_{ch}/d\eta \rangle$ in Pb+Pb collisions at $\sqrt{s_{NN}}$ = 2.76 and 5.02 TeV, Xe+Xe collisions at $\sqrt{s_{NN}}$ = 5.44 TeV, $p$+Pb collisions at $\sqrt{s_{NN}}$ = 5.02 TeV, and $p$+$p$ collisions at $\sqrt{s}$ = 5.02 and 13 TeV calculated using Eq. (\ref{eq:expansion}). 
}
    \label{thermo:expansion}
\end{figure}

The expansion coefficient or isobaric expansivity ($\alpha$) of a system is the measure of relative expansion per degree change in temperature at a constant pressure. For the hadronic medium at KFO, $\alpha$ is calculated using the following expression \cite{Sahu2021}: 

 \begin{equation}
 \label{eq:expansion}
\alpha = -\frac{\partial n/\partial T}{n}.
\end{equation} 

Figure \ref{thermo:expansion} shows $\alpha$ as a function $\langle dN_{ch}/d\eta \rangle$. The value of $\alpha$ is observed to be negative in the studied collision systems. $\alpha$ increases with increasing $\langle dN_{ch}/d\eta \rangle$. This is a consequence of the increasing particle density with $\langle dN_{ch}/d\eta \rangle$. For a denser system, the increase in temperature produces a small change in its volume, whereas for a less dense system, the change in volume is comparatively higher. 

\subsection{\label{sec:speed}Speed of sound}
\begin{figure}
    \centering
    \includegraphics[width=0.5\textwidth]{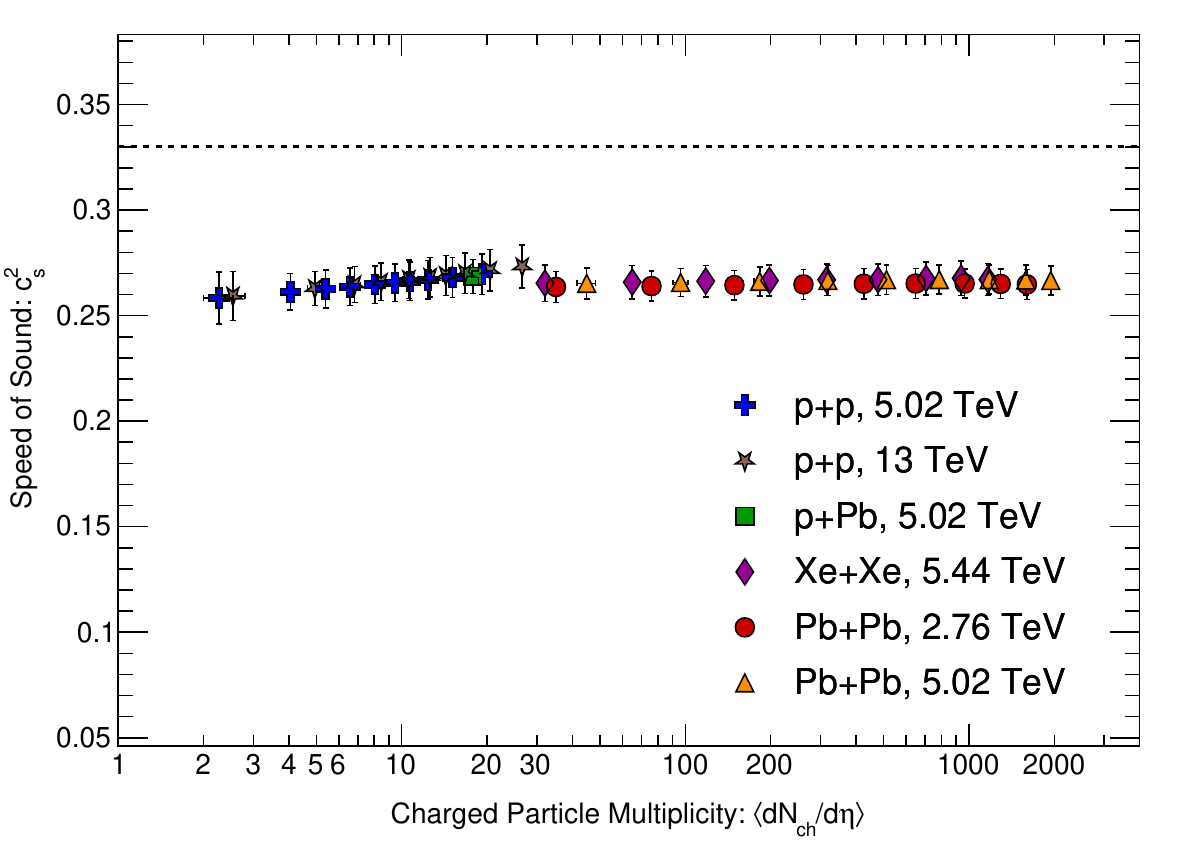}\hfill
    \caption{Squared speed of sound at kinetic freeze-out as a function of $\langle dN_{ch}/d\eta \rangle$ in Pb+Pb collisions at $\sqrt{s_{NN}}$ = 2.76 and 5.02 TeV, Xe+Xe collisions at $\sqrt{s_{NN}}$ = 5.44 TeV, $p$+Pb collisions at $\sqrt{s_{NN}}$ = 5.02 TeV, and $p$+$p$ collisions at $\sqrt{s}$ = 5.02 and 13 TeV calculated using Eq. (\ref{eq:sound}). The dotted line represents the value of $c_s^2$ for a non-interacting massless ideal gas. 
}
    \label{thermo:speed}
\end{figure}
The speed of sound in hadronic medium is closely related to its equation of state. The squared speed of sound is given as \cite{Khuntia2016,Jain2023}:
 \begin{equation}
    \label{eq:sound}
    c_s^2 = \left(\frac{\partial P}{\partial \epsilon}\right)_s = \left(\frac{\partial P/\partial T}{\partial \epsilon/\partial T}\right).
\end{equation} 
For a non-interacting massless ideal gas, the expected value of $c_s^2$ is 1/3 times the speed of light squared \cite{Ferrer2023}. Figure \ref{thermo:speed} shows $c_s^2$ plotted as a function of $\langle dN_{ch}/d\eta \rangle$. $c_s^2$ shows a weak dependence with $\langle dN_{ch}/d\eta \rangle$. We observe that for studied collision systems and energies, on an average $c_s^2$ = 0.266 $\pm$ 0.001.

\section{\label{sec:summary}Summary}
The charged hadron spectra measured by the ALICE Collaboration in Pb+Pb collisions at $\sqrt{s_{NN}}$ = 2.76 and 5.02 TeV, Xe+Xe collisions at $\sqrt{s_{NN}}$ = 5.44 TeV, $p$+Pb collisions at $\sqrt{s_{NN}}$ = 5.02 TeV, and $p$+$p$ collisions at $\sqrt{s}$ = 5.02 and 13 TeV have been studied in the framework of Tsallis statistics. We find that the Tsallis fits provide a better description of charged hadron spectra in peripheral Pb+Pb and Xe+Xe collisions than in more central collisions. In addition, we also observe an improvement in the fitting for lower multiplicity classes in $p$+$p$ collisions and a reasonably good description of the spectra in $p$+Pb collisions. Furthermore, the dependence of the fit parameters ($V$, $T$, and $q$) on the $\langle dN_{ch}/d\eta \rangle$ was also noted. It is observed that:
\begin{itemize}
\item Tsallis volume of the hadronic medium increases with increasing $\langle dN_{ch}/d\eta \rangle$, but the growth rate of volume in $p$+$p$ collisions is slower compared to Pb+Pb and Xe+Xe collisions, while $p$+Pb collisions show a similar trend to $p$+$p$ collisions. This suggests that small collision systems produce smaller hadronic systems at KFO for the same $\langle dN_{ch}/d\eta \rangle$ compared to large collision systems.
\item Temperature increases with increasing $\langle dN_{ch}/d\eta \rangle$ with a slow increase in small collision systems and a sharp rise in large collision systems.
\item The entropy index, $q$, increases with increasing $\langle dN_{ch}/d\eta \rangle$ in $p+p$ collisions, suggesting a deviation from equilibrium in high multiplicity collisions. $q$-parameter in $p$+Pb collisions is observed to follow the same trend as $p+p$ collisions at the corresponding $\langle dN_{ch}/d\eta \rangle$. However, for Pb+Pb and Xe+Xe collisions, $q$ decreases with increasing $\langle dN_{ch}/d\eta \rangle$, indicating that the system is closer to equilibrium in central collisions.
\end{itemize}
The thermodynamic properties of the hadronic medium in high-energy collisions, focusing on the dependence of energy density, pressure, particle density, entropy density, mean free path, Knudsen number, heat capacity, isothermal compressibility, expansion coefficient, and speed of sound on $\langle dN_{ch}/d\eta \rangle$ are studied. We find that:
\begin{itemize} 
\item Energy density increases with increasing $\langle dN_{ch}/d\eta \rangle$, with small collision systems having a higher energy density in the common $\langle dN_{ch}/d\eta \rangle$ region, possibly because of the smaller volumes of the hadronic medium compared to large collision systems. 
\item Pressure increases with increasing $\langle dN_{ch}/d\eta \rangle$, with small collision systems exhibiting a faster increase. This might be explained by the fact that less space for particles to move around in small collision systems results into a more rapid momentum exchange.
\item Number density increases with increasing $\langle dN_{ch}/d\eta \rangle$, with smaller systems exhibiting a faster increase owing to the smaller volume of the produced hadronic system.
\item Entropy density increases with increasing $\langle dN_{ch}/d\eta \rangle$, with smaller systems having a higher rate of increase.
\item The mean free path decreases with increasing $\langle dN_{ch}/d\eta \rangle$ due to the increased particle number density. Knudsen number was found to be much greater than unity, suggesting that produced hadronic medium is in the free streaming region.
\item Heat capacity increases with increasing $\langle dN_{ch}/d\eta \rangle$. This is a consequence of increase in particle number with increasing $\langle dN_{ch}/d\eta \rangle$. Increasing the temperature of a system by one unit will require more energy when the system contains a greater number of particles.
\item Isothermal compressibility decreases with increasing $\langle dN_{ch}/d\eta \rangle$ indicating near-ideal behavior with increasing multiplicity.
\item Expansion coefficient is negative for the studied collision systems and it increases with increasing $\langle dN_{ch}/d\eta \rangle$. This is due to the higher particle density associated with greater $\langle dN_{ch}/d\eta \rangle$, resulting in a smaller change in volume for the same amount of pressure in a relatively denser system. 
\item Squared speed of sound shows a weak dependence with $\langle dN_{ch}/d\eta \rangle$. We observe an average value of $c_s^2$ = 0.266 $\pm$ 0.001 for the studied collisions systems and energies. 
\end{itemize}

This study characterizes the thermodynamic properties of the produced hadronic system in different collisions across a range of LHC energies. 
We observe that high multiplicity $p$+$p$ collisions produce a hadronic medium with thermodynamic properties similar to peripheral heavy-ion collisions. These observations may contribute to the efforts to understand the properties of the medium formed in high-multiplicity $p$+$p$ collisions.

\begin{acknowledgments}
CJ acknowledges the financial support from DAE-DST Project No. 3015/I/2021/Gen/RD-I/13283.
\end{acknowledgments}

\bibliography{apssamp}

\begin{thebibliography}{44}%
\makeatletter
\providecommand \@ifxundefined [1]{%
 \@ifx{#1\undefined}
}%
\providecommand \@ifnum [1]{%
 \ifnum #1\expandafter \@firstoftwo
 \else \expandafter \@secondoftwo
 \fi
}%
\providecommand \@ifx [1]{%
 \ifx #1\expandafter \@firstoftwo
 \else \expandafter \@secondoftwo
 \fi
}%
\providecommand \natexlab [1]{#1}%
\providecommand \enquote  [1]{``#1''}%
\providecommand \bibnamefont  [1]{#1}%
\providecommand \bibfnamefont [1]{#1}%
\providecommand \citenamefont [1]{#1}%
\providecommand \href@noop [0]{\@secondoftwo}%
\providecommand \href [0]{\begingroup \@sanitize@url \@href}%
\providecommand \@href[1]{\@@startlink{#1}\@@href}%
\providecommand \@@href[1]{\endgroup#1\@@endlink}%
\providecommand \@sanitize@url [0]{\catcode `\\12\catcode `\$12\catcode `\&12\catcode `\#12\catcode `\^12\catcode `\_12\catcode `\%12\relax}%
\providecommand \@@startlink[1]{}%
\providecommand \@@endlink[0]{}%
\providecommand \url  [0]{\begingroup\@sanitize@url \@url }%
\providecommand \@url [1]{\endgroup\@href {#1}{\urlprefix }}%
\providecommand \urlprefix  [0]{URL }%
\providecommand \Eprint [0]{\href }%
\providecommand \doibase [0]{https://doi.org/}%
\providecommand \selectlanguage [0]{\@gobble}%
\providecommand \bibinfo  [0]{\@secondoftwo}%
\providecommand \bibfield  [0]{\@secondoftwo}%
\providecommand \translation [1]{[#1]}%
\providecommand \BibitemOpen [0]{}%
\providecommand \bibitemStop [0]{}%
\providecommand \bibitemNoStop [0]{.\EOS\space}%
\providecommand \EOS [0]{\spacefactor3000\relax}%
\providecommand \BibitemShut  [1]{\csname bibitem#1\endcsname}%
\let\auto@bib@innerbib\@empty
\bibitem [{\citenamefont {Ejiri}(2008)}]{PhysRevD.78.074507}%
  \BibitemOpen
  \bibfield  {author} {\bibinfo {author} {\bibfnamefont {S.}~\bibnamefont {Ejiri}},\ }\href {https://doi.org/10.1103/PhysRevD.78.074507} {\bibfield  {journal} {\bibinfo  {journal} {Phys. Rev. D}\ }\textbf {\bibinfo {volume} {78}},\ \bibinfo {pages} {074507} (\bibinfo {year} {2008})}\BibitemShut {NoStop}%
\bibitem [{\citenamefont {Aoki}\ \emph {et~al.}(2006)\citenamefont {Aoki}, \citenamefont {Endr{\H{o}}di}, \citenamefont {Fodor}, \citenamefont {Katz},\ and\ \citenamefont {Szab{\'{o}}}}]{Aoki2006}%
  \BibitemOpen
  \bibfield  {author} {\bibinfo {author} {\bibfnamefont {Y.}~\bibnamefont {Aoki}}, \bibinfo {author} {\bibfnamefont {G.}~\bibnamefont {Endr{\H{o}}di}}, \bibinfo {author} {\bibfnamefont {Z.}~\bibnamefont {Fodor}}, \bibinfo {author} {\bibfnamefont {S.~D.}\ \bibnamefont {Katz}},\ and\ \bibinfo {author} {\bibfnamefont {K.~K.}\ \bibnamefont {Szab{\'{o}}}},\ }\href {https://doi.org/10.1038/nature05120} {\bibfield  {journal} {\bibinfo  {journal} {Nature}\ }\textbf {\bibinfo {volume} {443}},\ \bibinfo {pages} {675} (\bibinfo {year} {2006})}\BibitemShut {NoStop}%
\bibitem [{\citenamefont {Adams}\ \emph {et~al.}(2005)\citenamefont {Adams} \emph {et~al.}}]{Adams2005}%
  \BibitemOpen
  \bibfield  {author} {\bibinfo {author} {\bibfnamefont {J.}~\bibnamefont {Adams}} \emph {et~al.} (\bibinfo {collaboration} {STAR Collaboration}),\ }\href {https://doi.org/10.1016/j.nuclphysa.2005.03.085} {\bibfield  {journal} {\bibinfo  {journal} {Nucl. Phys. A}\ }\textbf {\bibinfo {volume} {757}},\ \bibinfo {pages} {102} (\bibinfo {year} {2005})}\BibitemShut {NoStop}%
\bibitem [{\citenamefont {Adam}\ \emph {et~al.}(2017)\citenamefont {Adam} \emph {et~al.}}]{2017}%
  \BibitemOpen
  \bibfield  {author} {\bibinfo {author} {\bibfnamefont {J.}~\bibnamefont {Adam}} \emph {et~al.} (\bibinfo {collaboration} {ALICE Collaboration}),\ }\href {https://doi.org/10.1038/nphys4111} {\bibfield  {journal} {\bibinfo  {journal} {Nature Phys.}\ }\textbf {\bibinfo {volume} {13}},\ \bibinfo {pages} {535} (\bibinfo {year} {2017})}\BibitemShut {NoStop}%
\bibitem [{\citenamefont {Bjorken}\ \emph {et~al.}(2013)\citenamefont {Bjorken}, \citenamefont {Brodsky},\ and\ \citenamefont {Goldhaber}}]{Bjorken2013}%
  \BibitemOpen
  \bibfield  {author} {\bibinfo {author} {\bibfnamefont {J.~D.}\ \bibnamefont {Bjorken}}, \bibinfo {author} {\bibfnamefont {S.~J.}\ \bibnamefont {Brodsky}},\ and\ \bibinfo {author} {\bibfnamefont {A.~S.}\ \bibnamefont {Goldhaber}},\ }\href {https://doi.org/10.1016/j.physletb.2013.08.066} {\bibfield  {journal} {\bibinfo  {journal} {Phys. Lett. B}\ }\textbf {\bibinfo {volume} {726}},\ \bibinfo {pages} {344} (\bibinfo {year} {2013})}\BibitemShut {NoStop}%
\bibitem [{\citenamefont {Khachatryan}\ \emph {et~al.}(2017)\citenamefont {Khachatryan} \emph {et~al.}}]{Khachatryan2017}%
  \BibitemOpen
  \bibfield  {author} {\bibinfo {author} {\bibfnamefont {V.}~\bibnamefont {Khachatryan}} \emph {et~al.} (\bibinfo {collaboration} {CMS Collaboration}),\ }\href {https://doi.org/10.1016/j.physletb.2016.12.009} {\bibfield  {journal} {\bibinfo  {journal} {Phys. Lett. B}\ }\textbf {\bibinfo {volume} {765}},\ \bibinfo {pages} {193} (\bibinfo {year} {2017})}\BibitemShut {NoStop}%
\bibitem [{\citenamefont {Khuntia}\ \emph {et~al.}(2019{\natexlab{a}})\citenamefont {Khuntia}, \citenamefont {Sharma}, \citenamefont {Tiwari}, \citenamefont {Sahoo},\ and\ \citenamefont {Cleymans}}]{Khuntia2019}%
  \BibitemOpen
  \bibfield  {author} {\bibinfo {author} {\bibfnamefont {A.}~\bibnamefont {Khuntia}}, \bibinfo {author} {\bibfnamefont {H.}~\bibnamefont {Sharma}}, \bibinfo {author} {\bibfnamefont {S.~K.}\ \bibnamefont {Tiwari}}, \bibinfo {author} {\bibfnamefont {R.}~\bibnamefont {Sahoo}},\ and\ \bibinfo {author} {\bibfnamefont {J.}~\bibnamefont {Cleymans}},\ }\href {https://doi.org/10.1140/epja/i2019-12669-6} {\bibfield  {journal} {\bibinfo  {journal} {Eur. Phys. J. A}\ }\textbf {\bibinfo {volume} {55}} (\bibinfo {year} {2019}{\natexlab{a}})}\BibitemShut {NoStop}%
\bibitem [{\citenamefont {Azmi}\ \emph {et~al.}(2020)\citenamefont {Azmi}, \citenamefont {Bhattacharyya}, \citenamefont {Cleymans},\ and\ \citenamefont {Paradza}}]{Azmi2020}%
  \BibitemOpen
  \bibfield  {author} {\bibinfo {author} {\bibfnamefont {M.~D.}\ \bibnamefont {Azmi}}, \bibinfo {author} {\bibfnamefont {T.}~\bibnamefont {Bhattacharyya}}, \bibinfo {author} {\bibfnamefont {J.}~\bibnamefont {Cleymans}},\ and\ \bibinfo {author} {\bibfnamefont {M.}~\bibnamefont {Paradza}},\ }\href {https://doi.org/10.1088/1361-6471/ab6c33} {\bibfield  {journal} {\bibinfo  {journal} {J. Phys. G: Nucl. Part. Phys}\ }\textbf {\bibinfo {volume} {47}},\ \bibinfo {pages} {045001} (\bibinfo {year} {2020})}\BibitemShut {NoStop}%
\bibitem [{\citenamefont {Deb}\ \emph {et~al.}(2021)\citenamefont {Deb}, \citenamefont {Sarwar}, \citenamefont {Sahoo},\ and\ \citenamefont {e~Alam}}]{Deb2021}%
  \BibitemOpen
  \bibfield  {author} {\bibinfo {author} {\bibfnamefont {S.}~\bibnamefont {Deb}}, \bibinfo {author} {\bibfnamefont {G.}~\bibnamefont {Sarwar}}, \bibinfo {author} {\bibfnamefont {R.}~\bibnamefont {Sahoo}},\ and\ \bibinfo {author} {\bibfnamefont {J.}~\bibnamefont {e~Alam}},\ }\href {https://doi.org/10.1140/epja/s10050-021-00496-7} {\bibfield  {journal} {\bibinfo  {journal} {Eur. Phys. J. A}\ }\textbf {\bibinfo {volume} {57}} (\bibinfo {year} {2021})}\BibitemShut {NoStop}%
\bibitem [{\citenamefont {Sahu}\ and\ \citenamefont {Sahoo}(2021{\natexlab{a}})}]{Sahu2021}%
  \BibitemOpen
  \bibfield  {author} {\bibinfo {author} {\bibfnamefont {D.}~\bibnamefont {Sahu}}\ and\ \bibinfo {author} {\bibfnamefont {R.}~\bibnamefont {Sahoo}},\ }\href {https://doi.org/10.3390/physics3020016} {\bibfield  {journal} {\bibinfo  {journal} {Physics}\ }\textbf {\bibinfo {volume} {3}},\ \bibinfo {pages} {207} (\bibinfo {year} {2021}{\natexlab{a}})}\BibitemShut {NoStop}%
\bibitem [{\citenamefont {Khuntia}\ \emph {et~al.}(2016)\citenamefont {Khuntia}, \citenamefont {Sahoo}, \citenamefont {Garg}, \citenamefont {Sahoo},\ and\ \citenamefont {Cleymans}}]{Khuntia2016}%
  \BibitemOpen
  \bibfield  {author} {\bibinfo {author} {\bibfnamefont {A.}~\bibnamefont {Khuntia}}, \bibinfo {author} {\bibfnamefont {P.}~\bibnamefont {Sahoo}}, \bibinfo {author} {\bibfnamefont {P.}~\bibnamefont {Garg}}, \bibinfo {author} {\bibfnamefont {R.}~\bibnamefont {Sahoo}},\ and\ \bibinfo {author} {\bibfnamefont {J.}~\bibnamefont {Cleymans}},\ }\href {https://doi.org/10.1140/epja/i2016-16292-9} {\bibfield  {journal} {\bibinfo  {journal} {Eur. Phys. J. A}\ }\textbf {\bibinfo {volume} {52}} (\bibinfo {year} {2016})}\BibitemShut {NoStop}%
\bibitem [{\citenamefont {Jain}\ \emph {et~al.}(2023)\citenamefont {Jain}, \citenamefont {Gupta},\ and\ \citenamefont {Jena}}]{Jain2023}%
  \BibitemOpen
  \bibfield  {author} {\bibinfo {author} {\bibfnamefont {S.}~\bibnamefont {Jain}}, \bibinfo {author} {\bibfnamefont {R.}~\bibnamefont {Gupta}},\ and\ \bibinfo {author} {\bibfnamefont {S.}~\bibnamefont {Jena}},\ }\href {https://doi.org/10.3390/universe9040170} {\bibfield  {journal} {\bibinfo  {journal} {Universe}\ }\textbf {\bibinfo {volume} {9}},\ \bibinfo {pages} {170} (\bibinfo {year} {2023})}\BibitemShut {NoStop}%
\bibitem [{\citenamefont {Khuntia}\ \emph {et~al.}(2019{\natexlab{b}})\citenamefont {Khuntia}, \citenamefont {Tiwari}, \citenamefont {Sharma}, \citenamefont {Sahoo},\ and\ \citenamefont {Nayak}}]{PhysRevC.100.014910}%
  \BibitemOpen
  \bibfield  {author} {\bibinfo {author} {\bibfnamefont {A.}~\bibnamefont {Khuntia}}, \bibinfo {author} {\bibfnamefont {S.~K.}\ \bibnamefont {Tiwari}}, \bibinfo {author} {\bibfnamefont {P.}~\bibnamefont {Sharma}}, \bibinfo {author} {\bibfnamefont {R.}~\bibnamefont {Sahoo}},\ and\ \bibinfo {author} {\bibfnamefont {T.~K.}\ \bibnamefont {Nayak}},\ }\href {https://doi.org/10.1103/PhysRevC.100.014910} {\bibfield  {journal} {\bibinfo  {journal} {Phys. Rev. C}\ }\textbf {\bibinfo {volume} {100}},\ \bibinfo {pages} {014910} (\bibinfo {year} {2019}{\natexlab{b}})}\BibitemShut {NoStop}%
\bibitem [{\citenamefont {Sarkar}\ and\ \citenamefont {Ghosh}(2017)}]{PhysRevC.96.044901}%
  \BibitemOpen
  \bibfield  {author} {\bibinfo {author} {\bibfnamefont {N.}~\bibnamefont {Sarkar}}\ and\ \bibinfo {author} {\bibfnamefont {P.}~\bibnamefont {Ghosh}},\ }\href {https://doi.org/10.1103/PhysRevC.96.044901} {\bibfield  {journal} {\bibinfo  {journal} {Phys. Rev. C}\ }\textbf {\bibinfo {volume} {96}},\ \bibinfo {pages} {044901} (\bibinfo {year} {2017})}\BibitemShut {NoStop}%
\bibitem [{\citenamefont {Sarkar}\ \emph {et~al.}(2019)\citenamefont {Sarkar}, \citenamefont {Deb},\ and\ \citenamefont {Ghosh}}]{https://doi.org/10.48550/arxiv.1905.06532}%
  \BibitemOpen
  \bibfield  {author} {\bibinfo {author} {\bibfnamefont {N.}~\bibnamefont {Sarkar}}, \bibinfo {author} {\bibfnamefont {P.}~\bibnamefont {Deb}},\ and\ \bibinfo {author} {\bibfnamefont {P.}~\bibnamefont {Ghosh}},\ }\href {https://doi.org/10.48550/ARXIV.1905.06532} {\bibinfo {title} {arxiv:1905.06532v3}} (\bibinfo {year} {2019})\BibitemShut {NoStop}%
\bibitem [{\citenamefont {Tsallis}(1988)}]{Tsallis1988}%
  \BibitemOpen
  \bibfield  {author} {\bibinfo {author} {\bibfnamefont {C.}~\bibnamefont {Tsallis}},\ }\href {https://doi.org/10.1007/bf01016429} {\bibfield  {journal} {\bibinfo  {journal} {J. Stat. Phys.}\ }\textbf {\bibinfo {volume} {52}},\ \bibinfo {pages} {479} (\bibinfo {year} {1988})}\BibitemShut {NoStop}%
\bibitem [{\citenamefont {Cleymans}\ and\ \citenamefont {Worku}(2012{\natexlab{a}})}]{Cleymans2012}%
  \BibitemOpen
  \bibfield  {author} {\bibinfo {author} {\bibfnamefont {J.}~\bibnamefont {Cleymans}}\ and\ \bibinfo {author} {\bibfnamefont {D.}~\bibnamefont {Worku}},\ }\href {https://doi.org/10.1088/0954-3899/39/2/025006} {\bibfield  {journal} {\bibinfo  {journal} {J. Phys. G: Nucl. Part. Phys.}\ }\textbf {\bibinfo {volume} {39}},\ \bibinfo {pages} {025006} (\bibinfo {year} {2012}{\natexlab{a}})}\BibitemShut {NoStop}%
\bibitem [{\citenamefont {Hui}\ \emph {et~al.}(2018)\citenamefont {Hui}, \citenamefont {Jiang},\ and\ \citenamefont {Xu}}]{Hui2018}%
  \BibitemOpen
  \bibfield  {author} {\bibinfo {author} {\bibfnamefont {J.-Q.}\ \bibnamefont {Hui}}, \bibinfo {author} {\bibfnamefont {Z.-J.}\ \bibnamefont {Jiang}},\ and\ \bibinfo {author} {\bibfnamefont {D.-F.}\ \bibnamefont {Xu}},\ }\href {https://doi.org/10.1155/2018/7682325} {\bibfield  {journal} {\bibinfo  {journal} {Adv. High Energy Phys.}\ }\textbf {\bibinfo {volume} {2018}},\ \bibinfo {pages} {7682325} (\bibinfo {year} {2018})}\BibitemShut {NoStop}%
\bibitem [{\citenamefont {B{\'{\i}}r{\'{o}}}\ \emph {et~al.}(2017)\citenamefont {B{\'{\i}}r{\'{o}}}, \citenamefont {Barnaf\"{o}ldi}, \citenamefont {Bir{\'{o}}}, \citenamefont {\"{U}rm\"{o}ssy},\ and\ \citenamefont {Tak{\'{a}}cs}}]{Br2017}%
  \BibitemOpen
  \bibfield  {author} {\bibinfo {author} {\bibfnamefont {G.}~\bibnamefont {B{\'{\i}}r{\'{o}}}}, \bibinfo {author} {\bibfnamefont {G.}~\bibnamefont {Barnaf\"{o}ldi}}, \bibinfo {author} {\bibfnamefont {T.}~\bibnamefont {Bir{\'{o}}}}, \bibinfo {author} {\bibfnamefont {K.}~\bibnamefont {\"{U}rm\"{o}ssy}},\ and\ \bibinfo {author} {\bibfnamefont {{\'{A}}.}~\bibnamefont {Tak{\'{a}}cs}},\ }\href {https://doi.org/10.3390/e19030088} {\bibfield  {journal} {\bibinfo  {journal} {Entropy}\ }\textbf {\bibinfo {volume} {19}},\ \bibinfo {pages} {88} (\bibinfo {year} {2017})}\BibitemShut {NoStop}%
\bibitem [{\citenamefont {Si}\ \emph {et~al.}(2018)\citenamefont {Si}, \citenamefont {Li},\ and\ \citenamefont {Liu}}]{Si2018}%
  \BibitemOpen
  \bibfield  {author} {\bibinfo {author} {\bibfnamefont {R.-F.}\ \bibnamefont {Si}}, \bibinfo {author} {\bibfnamefont {H.-L.}\ \bibnamefont {Li}},\ and\ \bibinfo {author} {\bibfnamefont {F.-H.}\ \bibnamefont {Liu}},\ }\href {https://doi.org/10.1155/2018/7895967} {\bibfield  {journal} {\bibinfo  {journal} {Adv. High Energy Phys.}\ }\textbf {\bibinfo {volume} {2018}},\ \bibinfo {pages} {7895967} (\bibinfo {year} {2018})}\BibitemShut {NoStop}%
\bibitem [{\citenamefont {Lao}\ \emph {et~al.}(2017)\citenamefont {Lao}, \citenamefont {Liu},\ and\ \citenamefont {Lacey}}]{Lao2017}%
  \BibitemOpen
  \bibfield  {author} {\bibinfo {author} {\bibfnamefont {H.-L.}\ \bibnamefont {Lao}}, \bibinfo {author} {\bibfnamefont {F.-H.}\ \bibnamefont {Liu}},\ and\ \bibinfo {author} {\bibfnamefont {R.~A.}\ \bibnamefont {Lacey}},\ }\href {https://doi.org/10.1140/epja/i2017-12238-1} {\bibfield  {journal} {\bibinfo  {journal} {Eur. Phys. J. A}\ }\textbf {\bibinfo {volume} {53}} (\bibinfo {year} {2017})}\BibitemShut {NoStop}%
\bibitem [{\citenamefont {Cleymans}\ and\ \citenamefont {Worku}(2012{\natexlab{b}})}]{Cleymans20122}%
  \BibitemOpen
  \bibfield  {author} {\bibinfo {author} {\bibfnamefont {J.}~\bibnamefont {Cleymans}}\ and\ \bibinfo {author} {\bibfnamefont {D.}~\bibnamefont {Worku}},\ }\href {https://doi.org/10.1140/epja/i2012-12160-0} {\bibfield  {journal} {\bibinfo  {journal} {Eur. Phys. J. A}\ }\textbf {\bibinfo {volume} {48}} (\bibinfo {year} {2012}{\natexlab{b}})}\BibitemShut {NoStop}%
\bibitem [{\citenamefont {Bhattacharyya}\ \emph {et~al.}(2016)\citenamefont {Bhattacharyya}, \citenamefont {Cleymans},\ and\ \citenamefont {Mogliacci}}]{PhysRevD.94.094026}%
  \BibitemOpen
  \bibfield  {author} {\bibinfo {author} {\bibfnamefont {T.}~\bibnamefont {Bhattacharyya}}, \bibinfo {author} {\bibfnamefont {J.}~\bibnamefont {Cleymans}},\ and\ \bibinfo {author} {\bibfnamefont {S.}~\bibnamefont {Mogliacci}},\ }\href {https://doi.org/10.1103/PhysRevD.94.094026} {\bibfield  {journal} {\bibinfo  {journal} {Phys. Rev. D}\ }\textbf {\bibinfo {volume} {94}},\ \bibinfo {pages} {094026} (\bibinfo {year} {2016})}\BibitemShut {NoStop}%
\bibitem [{\citenamefont {Bhattacharyya}\ \emph {et~al.}(2017)\citenamefont {Bhattacharyya}, \citenamefont {Cleymans}, \citenamefont {Garg}, \citenamefont {Kumar}, \citenamefont {Mogliacci}, \citenamefont {Sahoo},\ and\ \citenamefont {Tripathy}}]{Bhattacharyya2017}%
  \BibitemOpen
  \bibfield  {author} {\bibinfo {author} {\bibfnamefont {T.}~\bibnamefont {Bhattacharyya}}, \bibinfo {author} {\bibfnamefont {J.}~\bibnamefont {Cleymans}}, \bibinfo {author} {\bibfnamefont {P.}~\bibnamefont {Garg}}, \bibinfo {author} {\bibfnamefont {P.}~\bibnamefont {Kumar}}, \bibinfo {author} {\bibfnamefont {S.}~\bibnamefont {Mogliacci}}, \bibinfo {author} {\bibfnamefont {R.}~\bibnamefont {Sahoo}},\ and\ \bibinfo {author} {\bibfnamefont {S.}~\bibnamefont {Tripathy}},\ }\href {https://doi.org/10.1088/1742-6596/878/1/012016} {\bibfield  {journal} {\bibinfo  {journal} {J. Phys.: Conf. Ser.}\ }\textbf {\bibinfo {volume} {878}},\ \bibinfo {pages} {012016} (\bibinfo {year} {2017})}\BibitemShut {NoStop}%
\bibitem [{\citenamefont {Cleymans}\ and\ \citenamefont {Paradza}(2020)}]{physics2040038}%
  \BibitemOpen
  \bibfield  {author} {\bibinfo {author} {\bibfnamefont {J.}~\bibnamefont {Cleymans}}\ and\ \bibinfo {author} {\bibfnamefont {M.~W.}\ \bibnamefont {Paradza}},\ }\href {https://doi.org/10.3390/physics2040038} {\bibfield  {journal} {\bibinfo  {journal} {Physics}\ }\textbf {\bibinfo {volume} {2}},\ \bibinfo {pages} {654} (\bibinfo {year} {2020})}\BibitemShut {NoStop}%
\bibitem [{\citenamefont {Acharya}\ \emph {et~al.}(2018)\citenamefont {Acharya} \emph {et~al.}}]{2018}%
  \BibitemOpen
  \bibfield  {author} {\bibinfo {author} {\bibfnamefont {S.}~\bibnamefont {Acharya}} \emph {et~al.} (\bibinfo {collaboration} {ALICE Collaboration}),\ }\href {https://doi.org/10.1007/jhep11(2018)013} {\bibfield  {journal} {\bibinfo  {journal} {J. High Energ. Phys.}\ }\textbf {\bibinfo {volume} {2018}},\ \bibinfo {pages} {13 (2018)}}\BibitemShut {NoStop}%
\bibitem [{\citenamefont {Acharya}\ \emph {et~al.}(2019{\natexlab{a}})\citenamefont {Acharya} \emph {et~al.}}]{Acharya2019}%
  \BibitemOpen
  \bibfield  {author} {\bibinfo {author} {\bibfnamefont {S.}~\bibnamefont {Acharya}} \emph {et~al.} (\bibinfo {collaboration} {ALICE Collaboration}),\ }\href {https://doi.org/10.1016/j.physletb.2018.10.052} {\bibfield  {journal} {\bibinfo  {journal} {Phys. Lett. B}\ }\textbf {\bibinfo {volume} {788}},\ \bibinfo {pages} {166} (\bibinfo {year} {2019}{\natexlab{a}})}\BibitemShut {NoStop}%
\bibitem [{\citenamefont {Acharya}\ \emph {et~al.}(2019{\natexlab{b}})\citenamefont {Acharya} \emph {et~al.}}]{Acharya20192}%
  \BibitemOpen
  \bibfield  {author} {\bibinfo {author} {\bibfnamefont {S.}~\bibnamefont {Acharya}} \emph {et~al.} (\bibinfo {collaboration} {ALICE Collaboration}),\ }\href {https://doi.org/10.1140/epjc/s10052-019-7350-y} {\bibfield  {journal} {\bibinfo  {journal} {Eur. Phys. J. C}\ }\textbf {\bibinfo {volume} {79}} (\bibinfo {year} {2019}{\natexlab{b}})}\BibitemShut {NoStop}%
\bibitem [{\citenamefont {Patra}\ \emph {et~al.}(2021)\citenamefont {Patra}, \citenamefont {Mohanty},\ and\ \citenamefont {Nayak}}]{NathPatra2021}%
  \BibitemOpen
  \bibfield  {author} {\bibinfo {author} {\bibfnamefont {R.~N.}\ \bibnamefont {Patra}}, \bibinfo {author} {\bibfnamefont {B.}~\bibnamefont {Mohanty}},\ and\ \bibinfo {author} {\bibfnamefont {T.~K.}\ \bibnamefont {Nayak}},\ }\href {https://doi.org/10.1140/epjp/s13360-021-01660-0} {\bibfield  {journal} {\bibinfo  {journal} {Eur. Phys. J. Plus}\ }\textbf {\bibinfo {volume} {136}} (\bibinfo {year} {2021})}\BibitemShut {NoStop}%
\bibitem [{\citenamefont {Abelev}\ \emph {et~al.}(2013)\citenamefont {Abelev} \emph {et~al.}}]{PhysRevC.88.044910}%
  \BibitemOpen
  \bibfield  {author} {\bibinfo {author} {\bibfnamefont {B.}~\bibnamefont {Abelev}} \emph {et~al.} (\bibinfo {collaboration} {ALICE Collaboration}),\ }\href {https://doi.org/10.1103/PhysRevC.88.044910} {\bibfield  {journal} {\bibinfo  {journal} {Phys. Rev. C}\ }\textbf {\bibinfo {volume} {88}},\ \bibinfo {pages} {044910} (\bibinfo {year} {2013})}\BibitemShut {NoStop}%
\bibitem [{\citenamefont {Acharya}\ \emph {et~al.}(2020)\citenamefont {Acharya} \emph {et~al.}}]{Acharya2020}%
  \BibitemOpen
  \bibfield  {author} {\bibinfo {author} {\bibfnamefont {S.}~\bibnamefont {Acharya}} \emph {et~al.} (\bibinfo {collaboration} {ALICE Collaboration}),\ }\href {https://doi.org/10.1016/j.physletb.2019.135043} {\bibfield  {journal} {\bibinfo  {journal} {Phys. Lett. B}\ }\textbf {\bibinfo {volume} {800}},\ \bibinfo {pages} {135043} (\bibinfo {year} {2020})}\BibitemShut {NoStop}%
\bibitem [{\citenamefont {Grosse-Oetringhaus}(2014)}]{GROSSEOETRINGHAUS201422}%
  \BibitemOpen
  \bibfield  {author} {\bibinfo {author} {\bibfnamefont {J.~F.}\ \bibnamefont {Grosse-Oetringhaus}},\ }\href {https://doi.org/https://doi.org/10.1016/j.nuclphysa.2014.10.003} {\bibfield  {journal} {\bibinfo  {journal} {Nuclear Physics A}\ }\textbf {\bibinfo {volume} {931}},\ \bibinfo {pages} {22} (\bibinfo {year} {2014})},\ \bibinfo {note} {qUARK MATTER 2014}\BibitemShut {NoStop}%
\bibitem [{\citenamefont {Liu}\ \emph {et~al.}(2023)\citenamefont {Liu}, \citenamefont {Yin},\ and\ \citenamefont {Zheng}}]{Liu2023}%
  \BibitemOpen
  \bibfield  {author} {\bibinfo {author} {\bibfnamefont {L.}~\bibnamefont {Liu}}, \bibinfo {author} {\bibfnamefont {Z.-B.}\ \bibnamefont {Yin}},\ and\ \bibinfo {author} {\bibfnamefont {L.}~\bibnamefont {Zheng}},\ }\href {https://doi.org/10.1088/1674-1137/aca38d} {\bibfield  {journal} {\bibinfo  {journal} {Chinese Phys. C}\ }\textbf {\bibinfo {volume} {47}},\ \bibinfo {pages} {024103} (\bibinfo {year} {2023})}\BibitemShut {NoStop}%
\bibitem [{\citenamefont {Loizides}\ \emph {et~al.}(2018)\citenamefont {Loizides}, \citenamefont {Kamin},\ and\ \citenamefont {d'Enterria}}]{PhysRevC.97.054910}%
  \BibitemOpen
  \bibfield  {author} {\bibinfo {author} {\bibfnamefont {C.}~\bibnamefont {Loizides}}, \bibinfo {author} {\bibfnamefont {J.}~\bibnamefont {Kamin}},\ and\ \bibinfo {author} {\bibfnamefont {D.}~\bibnamefont {d'Enterria}},\ }\href {https://doi.org/10.1103/PhysRevC.97.054910} {\bibfield  {journal} {\bibinfo  {journal} {Phys. Rev. C}\ }\textbf {\bibinfo {volume} {97}},\ \bibinfo {pages} {054910} (\bibinfo {year} {2018})}\BibitemShut {NoStop}%
\bibitem [{\citenamefont {Sahu}\ and\ \citenamefont {Sahoo}(2021{\natexlab{b}})}]{Sahu20212}%
  \BibitemOpen
  \bibfield  {author} {\bibinfo {author} {\bibfnamefont {D.}~\bibnamefont {Sahu}}\ and\ \bibinfo {author} {\bibfnamefont {R.}~\bibnamefont {Sahoo}},\ }\href {https://doi.org/10.1088/1361-6471/ac2cd6} {\bibfield  {journal} {\bibinfo  {journal} {J. Phys. G: Nucl. Part. Phys.}\ }\textbf {\bibinfo {volume} {48}},\ \bibinfo {pages} {125104} (\bibinfo {year} {2021}{\natexlab{b}})}\BibitemShut {NoStop}%
\bibitem [{\citenamefont {Shuryak}\ and\ \citenamefont {Zahed}(2013)}]{PhysRevC.88.044915}%
  \BibitemOpen
  \bibfield  {author} {\bibinfo {author} {\bibfnamefont {E.}~\bibnamefont {Shuryak}}\ and\ \bibinfo {author} {\bibfnamefont {I.}~\bibnamefont {Zahed}},\ }\href {https://doi.org/10.1103/PhysRevC.88.044915} {\bibfield  {journal} {\bibinfo  {journal} {Phys. Rev. C}\ }\textbf {\bibinfo {volume} {88}},\ \bibinfo {pages} {044915} (\bibinfo {year} {2013})}\BibitemShut {NoStop}%
\bibitem [{\citenamefont {Castorina}\ \emph {et~al.}(2020)\citenamefont {Castorina}, \citenamefont {Iorio}, \citenamefont {Lanteri}, \citenamefont {Satz},\ and\ \citenamefont {Spousta}}]{PhysRevC.101.054902}%
  \BibitemOpen
  \bibfield  {author} {\bibinfo {author} {\bibfnamefont {P.}~\bibnamefont {Castorina}}, \bibinfo {author} {\bibfnamefont {A.}~\bibnamefont {Iorio}}, \bibinfo {author} {\bibfnamefont {D.}~\bibnamefont {Lanteri}}, \bibinfo {author} {\bibfnamefont {H.}~\bibnamefont {Satz}},\ and\ \bibinfo {author} {\bibfnamefont {M.}~\bibnamefont {Spousta}},\ }\href {https://doi.org/10.1103/PhysRevC.101.054902} {\bibfield  {journal} {\bibinfo  {journal} {Phys. Rev. C}\ }\textbf {\bibinfo {volume} {101}},\ \bibinfo {pages} {054902} (\bibinfo {year} {2020})}\BibitemShut {NoStop}%
\bibitem [{\citenamefont {Mohanty}\ and\ \citenamefont {Alam}(2003)}]{PhysRevC.68.064903}%
  \BibitemOpen
  \bibfield  {author} {\bibinfo {author} {\bibfnamefont {B.}~\bibnamefont {Mohanty}}\ and\ \bibinfo {author} {\bibfnamefont {J.-e.}\ \bibnamefont {Alam}},\ }\href {https://doi.org/10.1103/PhysRevC.68.064903} {\bibfield  {journal} {\bibinfo  {journal} {Phys. Rev. C}\ }\textbf {\bibinfo {volume} {68}},\ \bibinfo {pages} {064903} (\bibinfo {year} {2003})}\BibitemShut {NoStop}%
\bibitem [{\citenamefont {Kadam}\ and\ \citenamefont {Mishra}(2015)}]{PhysRevC.92.035203}%
  \BibitemOpen
  \bibfield  {author} {\bibinfo {author} {\bibfnamefont {G.~P.}\ \bibnamefont {Kadam}}\ and\ \bibinfo {author} {\bibfnamefont {H.}~\bibnamefont {Mishra}},\ }\href {https://doi.org/10.1103/PhysRevC.92.035203} {\bibfield  {journal} {\bibinfo  {journal} {Phys. Rev. C}\ }\textbf {\bibinfo {volume} {92}},\ \bibinfo {pages} {035203} (\bibinfo {year} {2015})}\BibitemShut {NoStop}%
\bibitem [{\citenamefont {Bugaev}\ \emph {et~al.}(2013)\citenamefont {Bugaev}, \citenamefont {Oliinychenko}, \citenamefont {Sorin},\ and\ \citenamefont {Zinovjev}}]{Bugaev2013}%
  \BibitemOpen
  \bibfield  {author} {\bibinfo {author} {\bibfnamefont {K.~A.}\ \bibnamefont {Bugaev}}, \bibinfo {author} {\bibfnamefont {D.~R.}\ \bibnamefont {Oliinychenko}}, \bibinfo {author} {\bibfnamefont {A.~S.}\ \bibnamefont {Sorin}},\ and\ \bibinfo {author} {\bibfnamefont {G.~M.}\ \bibnamefont {Zinovjev}},\ }\href {https://doi.org/10.1140/epja/i2013-13030-y} {\bibfield  {journal} {\bibinfo  {journal} {Eur. Phys. J. A}\ }\textbf {\bibinfo {volume} {49}} (\bibinfo {year} {2013})}\BibitemShut {NoStop}%
\bibitem [{\citenamefont {Huovinen}\ and\ \citenamefont {Molnar}(2009)}]{PhysRevC.79.014906}%
  \BibitemOpen
  \bibfield  {author} {\bibinfo {author} {\bibfnamefont {P.}~\bibnamefont {Huovinen}}\ and\ \bibinfo {author} {\bibfnamefont {D.}~\bibnamefont {Molnar}},\ }\href {https://doi.org/10.1103/PhysRevC.79.014906} {\bibfield  {journal} {\bibinfo  {journal} {Phys. Rev. C}\ }\textbf {\bibinfo {volume} {79}},\ \bibinfo {pages} {014906} (\bibinfo {year} {2009})}\BibitemShut {NoStop}%
\bibitem [{\citenamefont {Bouras}\ \emph {et~al.}(2010)\citenamefont {Bouras}, \citenamefont {Moln\'ar}, \citenamefont {Niemi}, \citenamefont {Xu}, \citenamefont {El}, \citenamefont {Fochler}, \citenamefont {Greiner},\ and\ \citenamefont {Rischke}}]{PhysRevC.82.024910}%
  \BibitemOpen
  \bibfield  {author} {\bibinfo {author} {\bibfnamefont {I.}~\bibnamefont {Bouras}}, \bibinfo {author} {\bibfnamefont {E.}~\bibnamefont {Moln\'ar}}, \bibinfo {author} {\bibfnamefont {H.}~\bibnamefont {Niemi}}, \bibinfo {author} {\bibfnamefont {Z.}~\bibnamefont {Xu}}, \bibinfo {author} {\bibfnamefont {A.}~\bibnamefont {El}}, \bibinfo {author} {\bibfnamefont {O.}~\bibnamefont {Fochler}}, \bibinfo {author} {\bibfnamefont {C.}~\bibnamefont {Greiner}},\ and\ \bibinfo {author} {\bibfnamefont {D.~H.}\ \bibnamefont {Rischke}},\ }\href {https://doi.org/10.1103/PhysRevC.82.024910} {\bibfield  {journal} {\bibinfo  {journal} {Phys. Rev. C}\ }\textbf {\bibinfo {volume} {82}},\ \bibinfo {pages} {024910} (\bibinfo {year} {2010})}\BibitemShut {NoStop}%
\bibitem [{\citenamefont {Gombeaud}\ \emph {et~al.}(2009)\citenamefont {Gombeaud}, \citenamefont {Lappi},\ and\ \citenamefont {Ollitrault}}]{PhysRevC.79.054914}%
  \BibitemOpen
  \bibfield  {author} {\bibinfo {author} {\bibfnamefont {C.}~\bibnamefont {Gombeaud}}, \bibinfo {author} {\bibfnamefont {T.}~\bibnamefont {Lappi}},\ and\ \bibinfo {author} {\bibfnamefont {J.-Y.}\ \bibnamefont {Ollitrault}},\ }\href {https://doi.org/10.1103/PhysRevC.79.054914} {\bibfield  {journal} {\bibinfo  {journal} {Phys. Rev. C}\ }\textbf {\bibinfo {volume} {79}},\ \bibinfo {pages} {054914} (\bibinfo {year} {2009})}\BibitemShut {NoStop}%
\bibitem [{\citenamefont {Ferrer}\ and\ \citenamefont {Hackebill}(2023)}]{Ferrer2023}%
  \BibitemOpen
  \bibfield  {author} {\bibinfo {author} {\bibfnamefont {E.}~\bibnamefont {Ferrer}}\ and\ \bibinfo {author} {\bibfnamefont {A.}~\bibnamefont {Hackebill}},\ }\href {https://doi.org/10.1016/j.nuclphysa.2023.122608} {\bibfield  {journal} {\bibinfo  {journal} {Nuclear Physics A}\ }\textbf {\bibinfo {volume} {1031}},\ \bibinfo {pages} {122608} (\bibinfo {year} {2023})}\BibitemShut {NoStop}%
\end{thebibliography}%

\end{document}